\RequirePackage{snapshot}

\documentclass[12pt]{article}
\usepackage{amsmath}
\usepackage{graphicx,psfrag,epsf}
\usepackage{enumerate}
\usepackage{natbib}
\usepackage{url} %
\usepackage{xr}

\usepackage{amsmath}
\usepackage{amssymb}
\usepackage{bm}
\usepackage{dsfont}
\usepackage{pdflscape}

\usepackage{algpseudocode}
\usepackage{algorithm}

\usepackage{graphicx}

\usepackage{tikz}
\usetikzlibrary{arrows}
\usetikzlibrary{positioning}
\newdimen\nodeDist
\usepackage{amsthm}
\usepackage{thmtools}
\theoremstyle{definition}

\newtheorem{prop}{Proposition}[section]

\newcommand{\rr}{\kappa}
\newcommand{\iid}{\overset{iid}{\sim}} 
\newcommand{\approxsim}{\overset{approx}{\sim}} 
\newcommand{\poislik}[2]{\frac{\exp[-{#2}] [{#2}]^{#1} }{#1!}}
\newcommand{\x}{\mathbf{x}}
\newcommand{\pnb}{p_{\mathit{NB}}}
\newcommand{\pp}{p_{\mathit{P}}}

\newcommand{\ind}[1]{\mathds{1}(#1)}%

\newcommand{\f}{f}

\newcommand{\G}{Gamma}

\newcommand{\beq}{\begin{equation}}
\newcommand{\eeq}{\end{equation}}

\newcommand{\ben}{\begin{enumerate}}
\newcommand{\een}{\end{enumerate}}

\def\eqd{\,{\buildrel d \over =}\,}

\newcommand{\g}{g}
\newcommand{\M}{M}

\newcommand{\ghxi}[2]{g(\x_{#2}, T_{#1}, \M_{#1})}

\newcommand{\E}{\mathrm{E}}
\newcommand{\logit}{\mathrm{logit}}

\newcommand{\Var}{\mathrm{Var}}
\newcommand{\nc}{Z}
\def\eqd{\,{\buildrel d \over =}\,} 

\newcommand{\pc}{c}
\newcommand{\pd}{d}

\newcommand{\klein}{StAR}
\newcommand{\noth}{\f_{(h)}}

\newcommand{\loggamma}{\psi}

\newcommand{\blind}{0}

\algnewcommand\algorithmicinput{\textbf{Input:}}
\algnewcommand\INPUT{\item[\algorithmicinput]}

\algnewcommand\algorithmicoutput{\textbf{Output:}}
\algnewcommand\OUTPUT{\item[\algorithmicoutput]}

\begin{document}

\def\spacingset#1{\renewcommand{\baselinestretch}
{#1}\small\normalsize} \spacingset{1}

\if0\blind
{
  \title{\bf Log-Linear Bayesian Additive Regression Trees for Multinomial Logistic and Count Regression Models}
  \author{Jared S. Murray\thanks{
    The author gratefully acknowledges support from the National Science Foundation under grant numbers SES-1130706, SES-1631970, SES-1824555 and DMS-1043903. Any opinions, findings, and conclusions or recommendations expressed
in this material are those of the author(s) and do not necessarily reflect the views of the funding agencies. Thanks to P. Richard Hahn and Carlos Carvalho for helpful comments and suggestions on an early version of this manuscript.
    }\hspace{.2cm}\\
    University of Texas at Austin\\
    }
  \maketitle
} \fi

\if1\blind
{
  \bigskip
  \bigskip
  \bigskip
  \begin{center}
    {\LARGE\bf Log-Linear Bayesian Additive Regression Trees for Multinomial Logistic and Count Regression Models}
\end{center}
  \medskip
} \fi

\bigskip
\begin{abstract}
We introduce Bayesian additive regression trees (BART) for log-linear models including multinomial logistic regression and count regression with zero-inflation and overdispersion. BART has been applied to
nonparametric mean regression and binary classification problems in a range of
settings. However, existing applications of BART have been limited to models for Gaussian ``data'', either observed or latent. This is primarily because efficient MCMC algorithms are available for Gaussian likelihoods. But while many useful models are naturally cast in terms of latent Gaussian variables, many others are not -- including models considered in this paper.

We develop new data augmentation strategies and carefully specified prior distributions for these new models. Like the original BART prior, the new prior distributions are carefully constructed and calibrated to be flexible while guarding against overfitting. Together the new priors and data augmentation schemes allow us to implement an efficient MCMC sampler outside the context of Gaussian models. The utility of these new methods is illustrated with examples and an application to a previously published dataset.

\end{abstract}

\noindent%
\vfill

\newpage
\spacingset{1.45} %

\section{Introduction}

Since their introduction by \cite{Chipman2010}, Bayesian additive regression trees (BART) have been applied to nonparametric regression and classification problems in a wide range of settings. To date these have been limited to models for Gaussian data, perhaps after data augmentation (as in probit BART for binary classification). Although many useful models are naturally cast in terms of latent Gaussian variables, many others are not or have other, more convenient latent variable representations. This paper extends BART to a much wider range of models via a novel log-linear formulation that is easily incorporated into regression models for categorical and count responses. Adapting BART to the log-linear setting while maintaining the computational efficiency of the original BART MCMC algorithm requires careful consideration of prior distributions, one of the main contributions of this paper. 

{ The paper proceeds as follows: The remainder of this section reviews BART, including elements of the MCMC algorithm used for posterior inference. In Section \ref{sec:log-linear} we introduce new log-linear BART models for categorical and count responses. In Section \ref{sec:liks} we describe data augmentation and MCMC algorithms for these models. In Section \ref{sec:priors} we introduce new prior distributions and give details of posterior computation. In Section \ref{sec:examples} we present a large simulation study and an application to previously published data. In Section \ref{sec:conclusion} we conclude with discussion of extensions and areas for future work.  }

\subsection{Bayesian Additive Regression Trees (BART)}\label{sec:bart}
BART was introduced by \cite{Chipman2010} (henceforth CGM) as a nonparametric prior over a regression function $f(\cdot)$ designed to capture complex, nonlinear relationships and interactions. Our exposition in this section closely follows CGM. For observed data pairs  
$\{(y_i, \x_i); 1\leq i\leq n\}$ CGM consider the regression model

\begin{equation}
y_i = f(\x_i) + \epsilon_i,\quad \epsilon_i\iid N(0,\sigma^2)\label{eq:meanbart},
\end{equation}
where $\f$ is represented as the sum of many regression trees.

Each tree $T_h$ (for $1\leq h\leq m$) consists of a set of interior decision nodes with splitting rules of the form $x_{ij}<c$, and a set of $b_h$ terminal nodes. Each terminal node has an associated parameter, collected in the vector $\M_h = (\mu_{h1},\mu_{h2},\dots \mu_{hb_h})'$.  We use $T = \{T_h:1\leq h \leq m\}$ and $M = \{M_h:1\leq h \leq m\}$ to refer to the collections of trees/parameters.

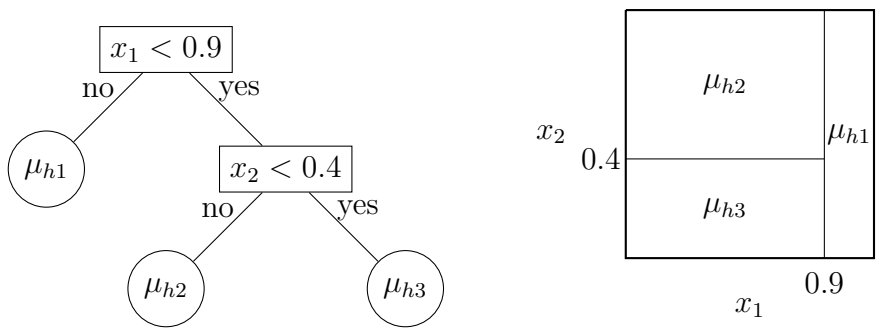
\begin{figure}[ht]
\begin{center}

\begin{tikzpicture}[
  scale=0.9,
    node/.style={%
      draw,
      rectangle,
    },
    node2/.style={%
      draw,
      circle,
    },
  ]

    \node [node] (A) {$x_1<0.9$};
    \path (A) ++(-135:\nodeDist) node [node2] (B) {$\mu_{h1}$};
    \path (A) ++(-45:\nodeDist) node [node] (C) {$x_2<0.4$};
    \path (C) ++(-135:\nodeDist) node [node2] (D) {$\mu_{h2}$};
    \path (C) ++(-45:\nodeDist) node [node2] (E) {$\mu_{h3}$};

    \draw (A) -- (B) node [left,pos=0.25] {no}(A);
    \draw (A) -- (C) node [right,pos=0.25] {yes}(A);
    \draw (C) -- (D) node [left,pos=0.25] {no}(A);
    \draw (C) -- (E) node [right,pos=0.25] {yes}(A);
\end{tikzpicture}
\hspace{0.05\linewidth}
\begin{tikzpicture}[scale=3.3]
\draw [thick, -] (0,1) -- (0,0) -- (1,0) -- (1,1)--(0,1);
\draw [thin, -] (0.8, 1) -- (0.8, 0);
\draw [thin, -] (0.0, 0.4) -- (0.8, 0.4);
\node at (-0.1,0.4) {0.4};
\node at (0.8,-0.1) {0.9};
\node at (0.5,-0.2) {$x_1$};
\node at (-0.3,0.5) {$x_2$};
\node at (0.9,0.5) {$\mu_{h1}$};
\node at (0.4,0.7) {$\mu_{h2}$};
\node at (0.4,0.2) {$\mu_{h3}$};
\end{tikzpicture}
\end{center}
\caption{(Left) An example binary tree, with internal nodes labeled by their splitting rules and terminal nodes labeled with the corresponding parameters $\mu_{ht}$ (Right) The corresponding partition of the sample space and the step function $g(\x, T_h, M_h)$.}
\label{fig:treestep}
\end{figure}

A tree and its associated decision rules induce a partition of the covariate space $\{\mathcal{A}_{h1},\dots,\mathcal{A}_{hb_h}\}$, where each element of the partition corresponds to a terminal node in the tree. Each pair $(T_h, M_h)$ parameterizes a step function $g$:
\beq
\g(\x, T_h, \M_h) = \mu_{ht}\text{ if }\x\in\mathcal{A}_{ht} \text{ (for $1\leq t\leq b_h$)}.\label{eq:partdef}
\eeq
An example tree and its corresponding step function are given in Figure \ref{fig:treestep}.
In BART a large number of these step functions are additively combined to obtain $f$:

\begin{equation}
f(\x) = \sum_{h=1}^m \g(\x, T_h, \M_h).\nonumber
\end{equation}

The prior on $(T_h, \M_h)$ strongly favors small trees and leaf parameters that are near zero (assuming the response variable is centered), constraining each term in the sum to be a ``weak learner''. Each tree is assigned an independent prior introduced by \cite{Chipman1998}, where trees are grown iteratively: Starting from the root node, the probability that a node at depth $d$ splits (is not terminal) is given by   
\begin{equation*}
\alpha (1+d)^{-\beta},\;\;\alpha\in (0,1),\;\beta\in [0,\infty).
\end{equation*}
 CGM propose $\alpha=0.95$ and $\beta=2$ as default values, which strongly favors small trees (of depth 2-3).
 A variable to split on is then selected uniformly at random, and given the selected variable a value to split at is selected according to a prior distribution defined over a grid.  If the $j^{th}$ variable is continuous the grid for variable $j$ is either uniformly spaced or given by a collection of observed quantiles of $\{x_{ij}: 1\leq i\leq n\}$. 
 For binary or ordinal variables, the cutpoints can be defined as the collection of all possible values. Unordered categorical variables with $q$ levels are generally expanded as $q$ binary variables indicating each level, although alternative coding schemes could be used instead. 

To set shrinkage priors on $M$ and avoid overfitting, CGM suggest scaling the data to lie in $\pm 0.5$ and assigning the leaf parameters independent priors:
\begin{equation}
\mu_{ht}\iid N(0,\sigma^2_\mu)\;\;\textrm{ where }\sigma_\mu=0.5/(k\sqrt{m}).\nonumber
\end{equation} 
CGM recommend $1\leq k\leq 3$, with $k=2$ as a reasonable default choice. This prior shrinks the individual basis functions strongly toward zero and yields a  $N(0, m\sigma^2_\mu)$ marginal prior for $\f(\x)$ at any covariate value. Since $\sqrt{m}\sigma_\mu=0.5/k$ this prior assigns approximately 95\% probability to the range of the transformed data ($\pm 0.5$) when $k=2$, so $\sigma_\mu$ (through $k$) can be used to calibrate the prior. %

\subsection{MCMC for BART: ``Bayesian backfitting''}\label{sec:bayesianbackfit}

A key ingredient in the MCMC sampler for BART is the ``Bayesian backfitting'' step, which we describe briefly here. (The term Bayesian backfitting is due to \cite{Hastie2000-oc}, who proposed a similar algorithm for MCMC sampling in additive models.) Let $T_{(h)}\equiv\{T_l: 1\leq l \leq m,\ l\neq h\}$ denote all but the $h^{th}$ tree with $M_{(h)}$ defined similarly. CGM's MCMC algorithm updates $(T_h, \M_h\mid T_{(h)}, M_{(h)}, -)$ in a block. This is simplified by the clever observation that
\beq
R_{hi} = \left( y_i - \sum_{l\neq h}^m \ghxi{l}{i}\right) \sim N(\ghxi{h}{i}, \sigma^2),\nonumber
\eeq
so that $(T_h, \M_h)$ only depends on the data through the vector of current partial residuals $R_h = (R_{h1}, R_{h2},\dots R_{hn})$. The partial residuals follow the Bayesian regression tree model described in \cite{Chipman1998}, so the 
Metropolis-Hastings update given there can be can be applied directly to sample from $(T_h, \M_h\mid -)$, treating $R_h$ as the observations. This delivers a proper sample from the correct conditional distributions.

Jointly updating the $(T_h, M_h)$ pairs in this way obviates the need for transdimensional MCMC algorithms (to cope with the fact that the length of $M_h$ changes with the depth of $T_h$), which can be delicate to construct \citep{Green1995-ey}. In addition, block updating parameters often accelerates the mixing of MCMC algorithms \citep{Liu1994-qm, Roberts1997-jg}. The efficiency of this blocked MCMC sampler is a key feature of BART, and one of the contributions of this paper is to generalize this sampler to a wider range of models where backfitting is infeasible.

\section{Log-linear BART Models}\label{sec:log-linear}

Extensions of the BART model in \eqref{eq:meanbart} have previously been limited to Gaussian models. CGM utilized BART for binary classification using a probit link and \cite{AlbertChib}'s data augmentation. \cite{Kindo2013} similarly extended BART to unordered categorical responses with latent Gaussian random variables in a multinomial probit regression model. \cite{Sparapani2016-cf} use a clever reparameterization to adapt probit BART to survival analysis. The focus on Gaussian models seems to be motivated by the desire to utilize the Bayesian backfitting MCMC algorithm.

However, many models either lack a natural representation in terms of observed or latent Gaussian random variables or have a different, more convenient latent variable formulation. We consider several such models below.
These models include one or more regression functions with positivity constraints. The natural extension of BART to this setting is obtained by expanding the log of the regression function into a sum of trees:
\begin{equation}
\log[\f(\x)] = \sum_{h=1}^m \g(\x, T_h, \M_h),\nonumber
\end{equation}
yielding log-linear Bayesian additive regression trees (that is, the log of the function is linear in the BART basis). We introduce log-linear BART models for categorical and count responses in the following subsections.

\subsection{Multinomial logistic regression models}\label{sec:mnl}

Suppose that for each realized value of the covariate vector $\x_i = (x_{i1},\dots,x_{ip})$ we observe $n_i$ observations falling into one of $1\leq j\leq c$ categories. Often $n_i=1$ for all $i$, as in the case with continuous covariates. Let $y_{ij}$ be the number of observations with covariate value $\x_i$ in category $j$ (so that $\sum_{j=1}^c y_{ij} = n_i$). We assume that the probability of observing category $j$ at a given covariate level is 
\beq
\pi_{j}(\x_i) = \frac{
  \f^{(j)}(\x_i)
}{
  \sum_{l=1}^c \f^{(l)}(\x_i)\nonumber
},
\eeq
or equivalently that the log odds in favor of category $j'$ over $j$ are given by
\beq
\log[\f^{(j')}(\x_i)] - \log[\f^{(j)}(\x_i)]\label{eq:unidlogodds}
\eeq
for any $j\neq j'$. 

We will assume that
$\log[\f^{(j)}(\x_i)] = \sum_{h=1}^m \g(\x, T^{(j)}_h, M^{(j)}_h),$
which induces a log-linear form for each of the log odds functions as defined in \eqref{eq:unidlogodds}. The result is a multinomial logistic BART model:
\beq
\pi_{j}(\x_i) = \frac{
  \exp\left[ \sum_{h=1}^m \g(\x, T_h^{(j)}, M_h^{(j)}) \right]
}{
  \sum_{l=1}^c \exp\left[ \sum_{h=1}^m \g(\x, T_h^{(l)}, M_h^{(l)}) \right]
}.\nonumber
\eeq
Here $T^{(j)}$ and  $M^{(j)}$ are trees and parameters governing each $f^{(j)}$.

As written this model is unidentified. Identification could be obtained by fixing some $\f^{(l)}(\cdot):= 1$, in which case $\f^{(l)}(\x)$ gives the odds of category $l$ against category $j$ at covariate value $\x$.  However, this prior depends on the arbitrary choice of a reference category. Instead, we use proper priors for each $\f^{(j)}$ and work in the unidentified space. This avoids asymmetries in the prior arising from the arbitrary choice of the reference category, and has some computational benefits as well (see Section~\ref{sec:lr-ex} in the supplemental material). Post-processing MCMC samples yields estimates of identified quantities like predicted probabilities or odds ratios.

\subsection{Count regression models, with overdispersion and zero-inflation}\label{sec:poissonmodel} 

For count responses we begin with Poisson or negative binomial models with mean function $\E(y_i\mid \x_i) = \mu_{0i}\f(\x_i)$. Here $\mu_{0i}$ is a fixed offset such as an adjustment for unit-level exposure, or we may take $\mu_{0i}\equiv \mu_0$ to center the prior for the regression function at $\mu_0$. We induce a log-linear model for the mean function by assuming
\begin{equation}
\log[\f(\x)] = \sum_{h=1}^m \g(\x, T_h, \M_h).\nonumber
\end{equation}
The Poisson model is completely specified by the mean function. 
The negative binomial regression model has an additional parameter $\rr$, 
which controls the degree of overdispersion relative to the Poisson. Under the negative binomial model,
\beq
\Var(y_i\mid \x_i) = \E(y_i\mid \x_i)\left(1 + \frac{\E(y_i\mid \x_i)}{\rr}\right).\nonumber
\eeq
As $\kappa\rightarrow\infty$, the negative binomial model converges to the Poisson. 
The probability mass function under the Poisson model is
\begin{equation}
\pp(y_i\mid \x_i, \f) = \poislik{y_i}{\mu_{0i}\f(\x_i)}\nonumber
\end{equation}
For the negative binomial model we have
\begin{equation}
\pnb(y_i\mid \x_i, \f, \rr) = \frac{\Gamma(\rr + y_i)}{\Gamma(\rr)y_i!}
\left(
\frac{
   \rr%
}{
  \rr + \mu_{0i}\f(\x_i)
}
\right)^\rr
\left(
\frac{
   \mu_{0i}\f(\x_i)%
}{
  \rr + \mu_{0i}\f(\x_i)
}
\right)^{y_i}.\nonumber
\end{equation}

Many datasets exhibit an excess of zero values. Zero inflated variants of Poisson or negative binomial regression models accommodate the extra zeros by adding a point mass component:
\begin{equation}
\Pr(Y_i=y_i\mid \x_i)=
  \begin{cases}
   (1-\omega(\x_i)) + \omega(\x_i)p(y_i\mid \x_i, \f, \kappa) & \text{if } y_i = 0 \\
   \omega(\x_i)p(y_i\mid \x_i, \f, \rr)     & \text{if } y_i > 0
  \end{cases}\nonumber
\end{equation}
where $p(y_i\mid \x_i, \f, \kappa)$ is the probability mass function of a Poisson or negative binomial with mean $\mu_{0i}\f(\x_i)$ and dispersion $\kappa$ and $1-\omega(\x_i)$ is the probability that a zero is due to the point mass component. 
We assume that 
\beq
\logit[1-\omega(\x_i)] = \log[1-\omega(\x_i)] - \log[\omega(\x_i)] \nonumber%
\eeq 
has a log-linear expansion, which will be induced through the redundant parameterization
\beq
\omega(\x_i) = \frac{\f^{(1)}(\x_i)}{\f^{(0)}(\x_i) + \f^{(1)}(\x_i)},\nonumber
\eeq
where $f^{(0)}$ and $f^{(1)}$ have independent log-linear BART priors as in the multinomial logistic regression model in Section~\ref{sec:mnl}.

\section{MCMC and Data Augmentation for Log-linear BART}\label{sec:liks}

Fitting the models in Section~\ref{sec:log-linear} is nontrivial: Some of the models lack a Gaussian representation, so CGM's Bayesian backfitting approach does not apply directly. 
However, the key element in CGM's MCMC sampler is actually a blocked MCMC update for each tree and its parameters, holding the other trees and parameters fixed.  CGM derive this update via Bayesian backfitting, but this is not strictly necessary. The general form of the update is summarized in Algorithm \ref{alg:backfit}, using notation defined below.

We have one or more functions that have a sum-of-trees representation on the log scale, so that $\log[\f(\x)] = \sum_{h=1}^m \g(\x, T_h, \M_h)$.
It will be convenient to work with $\f$ directly, so we define the following transformed parameters:
\begin{align}
\lambda_{ht} &= \exp(\mu_{ht}),\quad \Lambda_h = (\lambda_{h1},\lambda_{h2},\dots \lambda_{hb_h})',\nonumber
\end{align}
and note that 
$g(\x, T_h, \Lambda_h) = \exp[\g(\x, T_h, \M_h)]  = \lambda_{ht}\text{ if }\x\in\mathcal{A}_{ht} \text{ (for $1\leq t\leq b_h$)}$,
so
\beq
\f(\x) = \exp\left[\sum_{h=1}^m \g(\x, T_h, \M_h)\right] = \prod_{h=1}^m g(\x, T_h, \Lambda_h).\nonumber
\eeq
Additional parameters (such as $\kappa$ in the negative binomial regression model in Section~\ref{sec:poissonmodel}) or latent variables are collected in a vector $\theta$. In models with more than one regression function we consider MCMC updates for each regression function conditional on the others, which we also collect in $\theta$. 

\begin{algorithm} 
\caption{One step of the MCMC algorithm for updating a log-linear BART function parameterized by $T=\{T_h\}$ and $\Lambda=\{\Lambda_h\}$ ($1\leq h\leq m$)}
\label{alg:backfit}
\begin{algorithmic}
\INPUT{Data and current values for $T$, $\Lambda$, and other parameters/latent variables (in $\theta$)}
\OUTPUT{New values of $T$, $\Lambda$}
\For{$1\leq h \leq m$}\\
  \begin{enumerate}
  \item Propose $T^*_h\sim q(T^*_h;\ T_h)$
  \item Set $a\gets \frac{L(T^*_h;\ T_{(h)}, \Lambda_{(h)}, \theta, y)p(T^*_h)}{L(T_h;\ T_{(h)}, \Lambda_{(h)}, \theta, y)p(T_h)} \frac{q(T_h;\ T^*_h)}{q(T^*_h;\ T_h)}$
  \item Set  $T_h\gets T^*_h$ with probability $\min(1, a)$
  \item Sample $\Lambda_h\sim p(\Lambda_h\mid T_h, -)$
  \end{enumerate}
\EndFor
\end{algorithmic}
\end{algorithm}

Computing the conditional integrated likelihood function   
\beq
L(T_h;\ T_{(h)}, \Lambda_{(h)}, \theta, y) = \int L(T_h, \Lambda_h;\ T_{(h)}, \Lambda_{(h)}, \theta, y)p(\Lambda_h)d\Lambda_h\nonumber
\eeq
is a key step in Algorithm 1.  This is trivial in Gaussian BART models because CGM's normal prior is conjugate to the distribution of the observed or latent data.
Efficiently computing this integral under CGM's original prior in log-linear BART models is not as simple, since the prior is no longer conjugate. In particular, we will be concerned with likelihoods of the form
\begin{equation}
L(T, \Lambda; \Theta, y)  = \prod_{i=1}^n w_if(\x_i)^{u_i}\exp[v_i\f(\x_i)]\label{eq:poissontype}
\end{equation}
where $w_i$, $u_i$,and  $v_i$ are some functions of $\theta$ and $y_i$ that will vary depending on the model under consideration. To derive the corresponding conditional likelihood for $(T_h, \Lambda_h)$, define $\noth(\x) = \prod_{l\neq h} g(\x, T_l, \Lambda_l) $. This is the fit from all but the $h^{th}$ tree, and does not vary with $(T_h, \Lambda_h)$. Then we have
\begin{align}
L((T_h, \Lambda_h); T_{(h)}, \Lambda_{(h)}, y)&=\prod_{i=1}^n w_i f(\x_i)^{u_i}\exp[v_i\f(\x_i)]\nonumber\\
&=\prod_{i=1}^n w_i[\noth(\x_i)g(\x_i, T_h, \Lambda_h)]^{u_i}\exp[v_i\noth(\x_i)g(\x_i, T_h, \Lambda_h)]\nonumber\\
&=\prod_{t=1}^{b_h} 
\prod_{i: \x_i\in \mathcal{A}_{ht}} 
w_i
[\noth(\x_i)\lambda_{ht}]^{u_i}\exp[v_i\noth(\x_i)\lambda_{ht}]\label{eq:leaflik}\\
&= c_h \prod_{t=1}^{b_h} 
  \lambda_{ht}^{r_{ht}}
  \exp\left[- s_{ht}\lambda_{ht}\right],\nonumber
\end{align} 
where the outer product in \eqref{eq:leaflik} runs over the terminal nodes of $T_h$ and the inner product is over the observations with covariate values in the corresponding element of the partition (as defined in \eqref{eq:partdef}), and
\begin{align}
c_h = \prod_{i=1}^n w_i\noth(\x_i)^{u_i},\quad
r_{ht} = \sum_{i: \x_i\in A_{ht}} u_i,\quad
s_{ht} = \sum_{i: \x_i\in A_{ht}} \noth(\x_i)v_i,\nonumber
\end{align}
with $r_{ht}$ and $s_{ht}$ playing the role of conditional ``sufficient'' statistics.

To implement Algorithm 1, we need to compute the conditional integrated likelihood
\beq
L(T_h;\ T_{(h)}, \Lambda_{(h)}, y) = \int c_h \prod_{t=1}^{b_h} 
  \lambda_{ht}^{r_{ht}}
  \exp\left[- \lambda_{ht}s_{ht}\right] p(\Lambda_h) d\Lambda_h\label{eq:cilik}
\eeq
in step 2.
The original BART prior for $M_h$ induces independent lognormal priors for $\lambda_{ht}$, and the integral \eqref{eq:cilik} is unavailable under this prior. Before introducing a new conjugate prior in Section \ref{sec:priors}, we show how all the models in Section~\ref{sec:log-linear} admit simple data augmentation schemes that result in likelihood functions with multiple factors of the form \eqref{eq:poissontype}. This will allow us to use one blocked sampler to fit all the models in Section 2. 

\subsection{Data Augmentation for Multinomial Logistic Models}\label{sec:damultinom}

 The likelihood contribution for each distinct covariate value is
\begin{equation}
p_{MN}(y_i) =
\binom{n_i}{y_{i1}y_{i2}\dots y_{ic}}
\frac{
  \prod_{j=1}^c \f^{(j)}(\x_i)^{y_{ij}}
}{
  (\sum_{l=1}^c \f^{(l)}(\x_i))^{n_i}\label{eq:multi}
}.
\end{equation}

We augment the likelihood function by introducing a new latent variable $\phi_i$, and defining a joint model for $(\phi_i, y_i)$ where the marginal probability mass function of $y_i$ is \eqref{eq:multi} and $(\phi_i\mid y_i, -)\sim \G(n_i, \sum_{j=1}^c \f^{(j)}(\x_i))$ (recall that $n_i = \sum_{j=1}^c y_{ij}$). This yields the following augmented likelihood: 
\begin{align}
p_{MN}(y_i, \phi_i) &=
\binom{n_i}{y_{i1}y_{i2}\dots y_{ic}}
  \left(\prod_{j=1}^c \f^{(j)}(\x_i)^{y_{ij}}\right)\frac{\phi_i^{n_i-1}}{\Gamma(n_i)}\exp\left[-\phi_i\sum_{j=1}^c \f^{(j)}(\x_i)\right]\nonumber\\
&= 
\binom{n_i}{y_{i1}y_{i2}\dots y_{ic}}
\frac{\phi_i^{n_i-1}}{\Gamma(n_i)}\prod_{j=1}^c \f^{(j)}(\x_i)^{y_{ij}}\exp\left[-\phi_i \f^{(j)}(\x_i)\right].\label{eq:augmulti}
\end{align}
Note that given $\phi_i$ the augmented model \eqref{eq:augmulti} factors into separate terms for each $\f^{(j)}(\cdot)$, with each taking the form of \eqref{eq:poissontype}.

The ``gamma trick'' as a tool for dealing with sums or integrals in the denominator has appeared in other settings as well (e.g. \cite{Nieto-Barajas2004-kg,Walker2011-to, Caron2012-nd}). The same likelihood (up to an irrelevant constant) can also be derived via the Poisson-multinomial transformation \citep{Baker1994,Forster2010}, which adds an artificial Poisson distribution for the cell total $n_i$ parameterized by $\phi_i$ and $\sum_{j=1}^c \f^{(j)}(\x_i)$ (with a further prior on $\phi_i$, $p(\phi_i)\propto \phi_i^{-1}$). Since $n_i$ is often fixed by design, in our view casting the augmented model directly in terms of a proper joint probability model for $(y_i,\phi_i)$ is more transparent and removes any questions about the propriety of the posterior. 

Our data augmentation has some advantages over alternatives for logistic models: There is a single latent variable with a simple distribution for each distinct covariate value (not necessarily each observation). Additionally, the functions $f^{(j)}$ are conditionally independent given $\phi$ allowing for parallel updates to speed up the most computationally intensive step during MCMC. No other known augmentation for logistic models has all these features. In addition to proposing the current state-of-the-art Polya-Gamma data augmentation for logistic likelihoods, \cite{Polson2013-ym} give a recent review and comparison of several choices (including e.g. \cite{Holmes2006-ol,Fruhwirth-Schnatter2010-zz}). While these augmentations yield Gaussian models, they either require multiple latent variables per observation or latent variables with non-standard distributions. None yield conditional independence of the $f^{(j)}$'s.

In related work, \cite{Kindo2013} proposed a multinomial probit BART model using \cite{AlbertChib}'s data augmentation, which requires sampling from a truncated multivariate normal latent variable for each observation. It also requires the specification of a reference category and a prior for the covariance matrix over the latent Gaussian random variables, neither of which is easy or inconsequential (see \cite{burgette2010symmetric} for discussion about reference categories, and \cite{burgette2012trace} on covariance matrix priors in linear regression settings). It also does not result in conditional independence of the $f^{(j)}$'s.

\subsection{Data Augmentation for Count Models}\label{sec:dacount}

The Poisson model requires no data augmentation. The negative binomial and zero-inflated Poisson data augmentation schemes can be obtained via restrictions of the data augmentation for the zero-inflated negative binomial (ZINB) model, which we describe below. The likelihood contribution of a single observation under the ZINB model is

\begin{align}
\begin{split}
p_{ZINB}(y_i\mid \x_i, \f, f^{(0)}, f^{(1)}, \rr) =& 
\frac{f^{(1)}(\x_i)}{f^{(0)}(\x_i) + f^{(1)}(\x_i)}
\pnb(y_i\mid \x_i, \f, \rr)\\
&+\left(\frac{f^{(0)}(\x_i)}{f^{(0)}(\x_i) + f^{(1)}(\x_i)}\right){\ind{y_i=0}}\label{eq:likzinb}
\end{split}
\end{align}

Introducing $\xi_i,\phi_i\in (0, \infty)$ and $Z_i\in \{0,1\}$ we can define the data augmented likelihood:

\begin{align}
p_{ZINB}(y_i, Z_i, \phi_i, \xi_i \mid \f^{(0)}, \f_i, \kappa, \f)= 
&\f^{(0)}(\x_i)^{1-Z_i}\exp[-\phi_i\f^{(0)}(\x_i)]\label{eq:nbf1}\\
&\times\f^{(1)}(\x_i)^{Z_i}\exp[-\phi_i\f^{(1)}(\x_i)]\label{eq:nbf2}\\
&
   \times\f(\x_i)^{Z_iy_i}%
\exp\left[-Z_i\xi_i\mu_{0i}\f(\x_i)\right]\label{eq:nbf3}\\
&\times\left\{
\frac{1}{\Gamma(\rr)y_i!}
   \rr^\rr%
   \mu_{0i}^{y_i}%
\xi_i^{\rr+y_i-1}\exp\left[-\xi_i\rr\right]
\right\}^{Z_i}\\
&\times\ind{Z_i=1\text{ when }y_i>0} 
.\label{eq:zipaug}
\end{align}
The indicator in \eqref{eq:zipaug} enforces a support constraint on $Z_i$, which is a partially  latent variable indicating which component of the mixture generated the observation ($Z_i=0$ for observations assigned to the point-mass mixture component, and $Z_i=1$ for observations assigned to the non-degenerate count distribution). Since the nonzero responses must have come from the non-degenerate distribution, $Z_i$ is fixed at one when $y_i>0$.
\begin{prop}
\label{prop:zinbaug}
Integrating over $\xi_i, \phi_i,$ and $Z_i$ in 
\eqref{eq:nbf1}-\eqref{eq:zipaug}
yields \eqref{eq:likzinb}.
\end{prop} 
Note that given values for all the latent variables, the likelihood factors into terms of the form \eqref{eq:poissontype} for each of the log-linear functions (Eq. \eqref{eq:nbf1}-\eqref{eq:nbf3}). The augmented likelihood function for the negative binomial model {\em without} zero-inflation is obtained by fixing $Z_i=1$ for all $i$ and removing terms \eqref{eq:nbf1}, \eqref{eq:nbf2} and \eqref{eq:zipaug}. An augmented likelihood for the zero-inflated Poisson model is recovered by setting $\xi_i=1$ for all $i$ and dropping the remaining terms involving $\kappa$. Applying both restrictions leads to the Poisson likelihood function.

\section{Prior choice and posterior computation} \label{sec:priors}

Given the conditional likelihood 
\begin{equation}
L((T_h, \Lambda_h); T_{(h)}, \Lambda_{(h)}, \Theta, y) = c_h \prod_{t=1}^{b_h} 
  \lambda_{ht}^{r_{ht}}
  \exp\left[- \lambda_{ht}s_{ht}\right],\label{eq:condlikgen}
\end{equation}
from the previous section we would prefer a prior for $\lambda_{ht}$ that is 
\begin{enumerate}
\item Symmetric about $0$ on the log scale, since 
\beq
\log[\f(\x)] = \sum_{h=1}^m\log[g(\x, T_h, \Lambda_h) ] = \sum_{h=1}^m \sum_{t=1}^{b_h} \log(\lambda_{ht})\ind{\x\in \mathcal{A}_{ht}},\nonumber
\eeq 
so each tree contributes one $\log(\lambda_{ht})$ to the overall fit ($\log[\f(\x)]$) for any observation. Similar in spirit to the original BART model, each contribution should be relatively small and in either direction with equal prior probability.
\item Conjugate to \eqref{eq:condlikgen}, so we can compute the integrated likelihood \eqref{eq:cilik} in closed form and easily sample the terminal node parameters from their full conditional $p(\Lambda_h\mid T_h, -)$.
\end{enumerate}
Independent lognormal priors on $\lambda_{ht}$ satisfy 1, but not 2. Independent Gamma priors  satisfy 2, but not 1 - they are asymmetric on the log scale. Exact symmetry and conditional conjugacy requires a new prior, which we introduce below.

\subsection{A symmetric, conditionally conjugate prior}\label{sec:mixleaf}

Our strategy for deriving the new prior on $\lambda_{ht}$ is to ensure that in addition to symmetry and conjugacy, we have $\log[\f(\x)]\approxsim N(0, a^2_0)$ marginally at any covariate value $\x$. This allows us to use $a_0$ to calibrate the log-linear prior the same way that $\sigma_\mu$ parameter calibrates the original CGM prior. (Nonzero means for the log-linear regression function are handled via multiplicative offsets.) So with independent priors for $\lambda_{ht}$, we require that $\E(\log[\lambda_{ht}])=0$ and $\Var(\log[\lambda_{ht}])=a_0^2/m$. Typically $m$ is large, so the normal approximation to the marginal distribution of $\log[\f(\x)]$ will be accurate by the central limit theorem. The specific prior below is somewhat complex, but the end result is very similar to CGM's leaf prior and has a single, interpretable tuning parameter (for a fixed $m$).

Our proposed leaf prior is a mixture of generalized inverse Gaussian (GIG) distributions. GIG distributions are characterized by their density function
\beq
p_{GIG}(\lambda\mid \eta, \chi, \psi) = 
\frac{\lambda^{\eta-1}
\exp\left[{-\frac{1}{2}} \left( \chi/\lambda + \psi\lambda \right) \right]}{\nc(\eta, \chi, \psi)},\nonumber
\eeq
with normalizing constant 
\beq
\nc(\eta, \chi, \psi) =
\begin{cases}
\Gamma(\eta)\left(\frac{2}{\psi}\right)^{\eta} 
  & \text{if } \eta>0,\ \chi=0,\ \psi > 0\\
\Gamma(-\eta)\left(\frac{2}{\chi}\right)^{-\eta}   
  & \text{if } \eta<0,\ \chi>0,\ \psi = 0\\
\frac{2K_\eta(\sqrt{\psi\chi})}{(\psi/\chi)^{(\eta/2)}} 
  & \text{if } \chi > 0,\ \psi>0,\nonumber
\end{cases}
\eeq
where $K_\eta(x)$ is the modified Bessel function of the second kind.
The gamma and inverse gamma distributions are recovered when $\chi=0$ and $\psi=0$, respectively. This distribution is also conjugate to \eqref{eq:condlikgen}. Our mixture prior is given by 
\begin{align}
p_\lambda(\lambda_{ht}\mid \pc,\pd) %
=\frac{1}{2}
p_{GIG}(\lambda_{ht}\mid -\pc, 2\pd, 0)
+\frac{1}{2}p_{GIG}(\lambda_{ht}\mid \pc, 0, 2\pd).
\nonumber
\end{align}
where $\pc$ and $\pd$ are parameters that will be determined by $a_0$. As a mixture of GIG distributions this prior is also conjugate to \eqref{eq:condlikgen}. We refer to this as the $P_\lambda(\pc, \pd)$ distribution.

The $P_{\lambda}(c,d)$ distribution has the following simple stochastic representation:
\begin{align*}
W_{ht}&\sim Gamma(\pc, \pd)\\
U_{ht}&\sim Bernoulli(1/2)\\
\lambda_{ht} &= U_{ht}W_{ht} + (1-U_{ht})(1/W_{ht}),
\end{align*}
By construction the implied prior on $\mu_{ht}$ is symmetric about $0$ since $\mu_{ht} = \log(W_{ht})$  or $-\log(W_{ht})$ with equal probability. (The $W$ and $U$ random variables are never instantiated and only introduced here for exposition.)

The parameters $\pc,\pd$ can be set from user-supplied values of $a_0$ and $m$. The optimal values are not available in closed form (although they are easy to obtain numerically) but for a large number of trees and/or a small value of $a_0$, the values of $\pc$, $\pd$ also have simple approximate values. These results are summarized in Propositions~\ref{prop:pars} and~\ref{prop:parsapprox}.  
\begin{prop}
\label{prop:pars}
If $\lambda\sim P_\lambda(\pc, \pd)$, then $\Var(\lambda) = a_0^2/m$ when $\loggamma''(\pc) = a_0^2/m$ and $\pd = \exp(\loggamma'(\pc))$. Here $\loggamma(\pc) = \log[\Gamma(\pc)]$, and $\loggamma'$ and $\loggamma''$ are its first and second derivatives. The function $\loggamma''(\pc)$ is monotone decreasing and hence invertible on $\mathbb{R}^+$, so the solutions to these equations are unique.
\end{prop}
\begin{prop}
\label{prop:parsapprox}
For small values of $a_0^2/m$, the values of $\pc$ and $\pd$ from Proposition~\ref{prop:pars} are approximately $\pc\approx m/a_0^2 +0.5$ and $\pd\approx m/a_0^2$.
\end{prop}

\begin{figure}[ht]
\begin{center}
{ \includegraphics[width=0.8\linewidth]{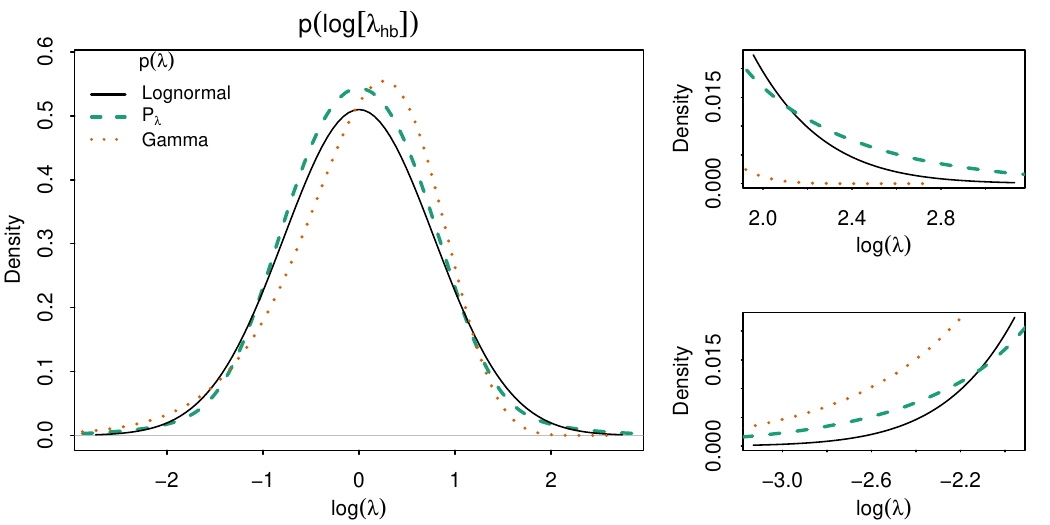} 
}
\end{center}
\caption{The proposed $P_\lambda$ node-parameter prior (green dashed line) compared to CGM's log-normal prior on $\lambda_{ht}$ (solid black) and a Gamma prior calibrated to have the same moments on the log scale (dashed orange). Here $m=25$ and $a_0=3.5/\sqrt{2}$}.
\label{fig:nodeprior}
\end{figure}

One could calibrate a gamma prior similarly, and in fact the shape and rate parameters will be the same as $\pc$ and $\pd$ in Proposition~\ref{prop:pars} (respectively). Figure~\ref{fig:nodeprior} compares the calibrated $P_\lambda$ and log-gamma priors to CGM's normal priors for $m=25$ and $a_0=3.5/\sqrt{2}$, which are actual parameter settings we will use later. The log-gamma prior is asymmetric, compared to the log-$P_\lambda$ prior which is symmetric and has slightly heavier tails than the normal. The log-gamma and log-$P_\lambda$ priors both become increasingly close to the normal distribution as $a_0^2/m\rightarrow 0$, but the asymmetry in the log-gamma prior for small values of $m$ is undesirable. The $P_\lambda$ prior is a more reasonable default choice for the entire range of $a_0$ and $m$ values.

\subsection{Posterior computation} \label{sec:mcmc}

With the prior specified we can now fill in the details of steps 1-4 in Algorithm \ref{alg:backfit}:

\begin{enumerate}
\item[Steps 1-3.] 
We utilize the grow, prune, change and swap proposal moves described by CGM (originally introduced in \cite{Chipman1998}) but any proposals could be used (see e.g. \cite{Denison1998-hl,Wu2007,Pratola2013} for other possibilities). The integrated likelihood function that appears in the  acceptance ratio is
\begin{align}
L(T_h; T_{(h)}, \Lambda_{(h)}, \Theta, y) &= c_h \prod_{t=1}^{b_h} \int
  \lambda_{ht}^{r_{ht}}
  \exp\left[- \lambda_{ht}s_{ht}\right] p_\lambda(\lambda_{ht}\mid \pc, \pd) d\lambda_{ht}\nonumber\\
  &=c_h \prod_{t=1}^{b_h}\frac{Z(-\pc + r_{ht}, 2\pd, 2s_{ht}) + Z(\pc + r_{ht}, 0, 2[\pd+s_{ht}])}
         {2Z(\pc, 0, 2\pd)}\label{eq:intlikexp}
\end{align}
using the fact that $Z(\pc, 0, 2\pd) = Z(-\pc, 2\pd, 0)$. The leading term $c_h$ cancels in the Metropolis-Hastings acceptance ratio, but the denominator in \eqref{eq:intlikexp} does not when the proposal changes the dimension of the partition (e.g. grow/prune moves).
\item[Step 4.] Sample $(\Lambda_h\mid T_h, T_{(h)}, \Lambda_{(h)})$ from its full conditional. The components of $\Lambda_h$ are conditionally independent with full conditional distributions
\begin{align}
p(\lambda_{ht}\mid -)
\propto\ 
&\lambda_{ht}^{(-\pc + r_{ht})}
\exp\left[{-\frac{1}{2}} \left( 2\pd/\lambda_{ht} + 2s_{ht}\lambda_{ht} \right) \right]
&+\lambda_{ht}^{(\pc + r_{ht})}
\exp\left[{-\frac{1}{2}} \left( 2\pd+ 2s_{ht} \right)\lambda_{ht} \right].\nonumber
\end{align}
This distribution is a mixture of GIG distributions:
\beq
p(\lambda_{ht}\mid -) = \pi_{ht}p_{GIG}(-\pc + r_{ht}, 2\pd, 2s_{ht}) + (1-\pi_{ht})p_{GIG}(\pc + r_{ht}, 0, 2[\pd+s_{ht}])\nonumber
\eeq
where
\beq
\pi_{ht} = \frac{Z(-\pc + r_{ht}, 2\pd, 2s_{ht})}{Z(-\pc + r_{ht}, 2\pd, 2s_{ht}) + Z(\pc + r_{ht}, 0, 2[\pd+s_{ht}])}.\nonumber
\eeq
\end{enumerate}

Algorithm 1 forms the backbone of MCMC in log-linear BART models, with additional parameters or latent variables sampled from their conditional distributions in further MCMC steps. In the following subsections we describe how to calibrate the $P_\lambda$ prior for the models in Section 2 and outline posterior sampling. 

\subsection{Prior choice and posterior computation for multinomial logistic models}

In the multinomial logistic BART model, for any two outcome categories $j\neq j'$ the log odds in favor of $j'$ are given by
\beq
\log[\f^{(j')}(\x_i)] - \log[\f^{(j)}(\x_i)],\label{eq:unidlogodds2}
\eeq
and each function $\f^{(l)}(\cdot)$ has an independent log-linear BART prior parameterized by $(T^{(l)}, \Lambda^{(l)})$ (for $1\leq l\leq c$). We assume that the prior on each $f^{(l)}(\cdot)$ uses the same number of trees $m$ and parameter $a_0$ in the $P_\lambda$ prior. Then the induced prior on \eqref{eq:unidlogodds2} is approximately $N(0, 2a_0^2)$, so $a_0$ can be chosen to reflect prior beliefs about the plausible range of the log odds functions.  Since the log-odds lie within $(-2\sqrt{2}a_0, 2\sqrt{2}a_0)$ at any covariate value with probability approximately 0.95 under the prior, $a_0 = 3.5/\sqrt{2}$ is a reasonable default choice.

A single step of the MCMC sampler proceeds as follows:
\begin{enumerate}
\item For $1\leq i \leq n$, draw $\phi_i\sim \G(n_i, \sum_{j=1}^c \f^{(j)}(\x_i))$. This is a direct consequence of the data augmentation, which was conditional on $y_i$ and the regression functions.
\item For $1\leq j \leq c$, \emph{independently} update the parameters of $\f^{(j)}$ using Algorithm \ref{alg:backfit} and the expressions in Section \ref{sec:mcmc} with
\begin{equation*}
r_{ht} = \sum_{i: \x_i\in A^{(j)}_{ht}} y_{ij},\quad 
s_{ht} = \sum_{i: \x_i\in A^{(j)}_{ht}} \phi_{i} \noth^{(j)}(\x_i)
\end{equation*}
where $\noth^{(j)}(\x_i) = \prod_{l\neq h} g(\x, T^{(j)}_h, \Lambda^{(j)}_h)$ is the fit from all but the $h^{th}$ tree.
\end{enumerate}

The augmentation in  \eqref{eq:augmulti} yields a very convenient MCMC algorithm: There is a \emph{single} augmented variable for each covariate value, regardless of the number of categories or observations, and it has a standard, untruncated distribution. Further, the $c$ regression functions are conditionally independent given the latent variable. %

\subsection{Prior choice and posterior computation for count models}\label{sec:prior-count}

We describe prior specification and MCMC sampling for the most complex case, the zero-inflated negative binomial. Prior specification is similar in negative binomial or zero-inflated Poisson models.
Specializations of the MCMC algorithm to the negative binomial or zero-inflated Poisson follow 
from the discussion at the end of Section \ref{sec:dacount}. 

Recall that the probability of observing an ``excess'' zero is
\beq
1-\omega(\x_i) = \frac{\f^{(0)}(\x_i)}{\f^{(0)}(\x_i) + \f^{(1)}(\x_i)}.\nonumber
\eeq
Similar to the previous subsection, independent log-linear BART priors on $f^{(0)}$ and $f^{(1)}$ with common values of the concentration parameter and number of trees (say $a_{z0}$ and $m_z$) induce a log-linear BART logistic regression model:
\beq
\logit[1-\omega(\x)] = \log[\f^{(0)}(\x)] - \log[\f^{(1)}(\x)] %
\label{eq:logistic-omega}
\eeq
The log-odds of observing an excess zero at any covariate value \eqref{eq:logistic-omega} is approximately distributed $N\left(0, 2a_{z0}^2\right)$ marginally, so $a_{z0}$ may be chosen based on plausible values for the odds function.\footnote{As pointed out by a reviewer, in some contexts it may be desirable to shrink toward some particular value for $1-\omega(\x)$; this can be accomplished by setting $1-\omega(\x_i) = \frac{n_0\f^{(0)}(\x_i)}{n_0\f^{(0)}(\x_i) + n_1\f^{(1)}(\x_i)}$, which centers the prior at $n_0/(n_0+n_1)$ with increasing values of $n_0+n_1$ imply stronger shrinkage.} As defaults we suggest $m_z=100$ and $a_0 = 3.5/\sqrt{2}$. 

In the zero-inflated model, $\mu_{0i}\f(\x_i)$ is the mean of the non-point mass component of the zero-inflated model and $\f(\cdot)$ has a log-linear BART prior with $m$ trees and concentration parameter $a_0$. Assuming $\mu_{0i}=\mu_0$, a reasonable default prior is obtained by positing a near-maximum value (or upper quantile of the empirical distribution) of $y$, say $y^*$, and setting $a_0 = 0.5[\log(y^*)-\log(\mu_0)]$. Then $\Pr(\f(\x_i)\leq y^*) \approx 0.975$ marginally, since $\log[\f(\x_i)]\approxsim N(0, a_0^2)$. For large values of $\mu_0$ it may be necessary to inflate this value to cover plausible low values for $\mu$.%
For $\kappa$, we use beta prime priors:  $p(\kappa)\propto \kappa^{a_\kappa-1}(1+\kappa)^{-a_\kappa+b_\kappa}$. This is a heavy-tailed prior which is equivalent to a $Beta(a_\kappa, b_\kappa)$ prior on $\kappa/(1+\kappa)$. Gamma priors are another reasonable choice (e.g. \cite{zhou2012lognormal}). %

Posterior sampling for the ZINB model has many more steps than the multinomial logistic regression model, and is outlined in Section~\ref{sec:zinb-mcmc} of the supplemental material. The primary innovation is three applications of Algorithm~\ref{alg:backfit} that can be run in parallel, with all the remaining parameters updated in a single block for efficiency.

\section{Illustrations and applications}\label{sec:examples}

\subsection{Simulation: Multinomial Logistic Regression}

We compared default and cross-validated multinomial logistic BART models (BART-default and BART-CV, respectively) with several classification methods using 20 datasets taken from the UCI repository and processed as in \cite{doweneed}. The primary purpose of this exercise is to establish multinomial logistic BART as having reasonable classification performance. We do not expect BART to necessarily outperform other machine learning methods designed and tuned for classification accuracy, but if BART can be established as a plausible classifier then we have some license to include log-linear BART priors as building blocks in more complicated Bayesian models where cross-validation is complicated or infeasible. To this end we also compared the performance of default and cross-validated BART models. Default variants require less computation and yield valid posterior inference, which may be desirable in their own right, but are essential if we have complex models with multiple nonparametric regression functions.

We chose to include all the datasets with 3-6 outcome categories and between 100 and 3,000 observations. Each dataset was randomly split into training and validation sets (comprising 80\% and 20\% of the data, respectively). For methods using cross-validation we performed 10-fold CV using the training set to choose parameter settings with the best estimated accuracy, refit to the entire training set using the selected parameters, and then evaluated performance on the held-out validation set. We repeated this procedure ten times, yielding ten estimates of out of sample accuracy per dataset and method.

We consider two potential variants of BART-default: one that sets the number of trees per category to 100, so that the log-odds functions involve $200$ trees, and one that sets the number of trees per category such that the {\em total} number of trees is as close to 200 as possible. Both set $a_0 = 3.5/\sqrt{2}$.
BART-CV was evaluated over range of $m$ that included both default rules for the number of trees (approximately 200 total, or 100 per each outcome category) and $25$ trees per category\footnote{As noted by a referee, the normal approximation used to calibrate the prior might not hold as well with few trees -- the symmetry and slightly heavier tails of the $P_\lambda$ generate a prior that puts somewhat less mass in the central interval. Given how close the $P_\lambda$ prior is to normal, we expect this prior is reasonable in any event.}. Possible values for $a_0$ included $2/\sqrt{2}, 3.5/\sqrt{2}$ (the default choice) and $6/\sqrt{2}$. Other methods included random forests, gradient boosted models, penalized multinomial probit regression, a support vector machine using radial basis functions, and a single layer neural net\footnote{We attempted to include \cite{Kindo2013}'s multinomial probit BART, but the accompanying R package routinely crashed during simulations. We expect that it would perform similar to multinomial logistic BART in cross-validation, at substantially increased computational cost due to the need to update several latent Gaussian variables per covariate value as well as a latent covariance matrix, and to cross-validate the choice of reference category in addition to $m$ and the parameters of the covariance matrix prior. (\cite{Kindo2013} propose no default settings for reference category or prior on the covariance matrix.)}. Each method was evaluated over its default parameter grid in the R package \texttt{caret} \citep{JSSv028i05,caret-package}.

\begin{table}[ht] 
\centering

\resizebox{6.5in}{!}{

\begin{tabular}{l|l|l|l|l|l|l}
\hline
  & rf & gbm & mno & svm & nnet & bart-cv\\
\hline
balance-scale & \textit{\color{gray}0.845 (0.021)} & \textit{\color{gray}0.92 (0.006)} & \textit{\color{gray}0.897 (0.018)} & \textit{\color{gray}0.91 (0.022)} & \textit{0.967 (0.017)}* & {0.932 (0.007)}\\
\hline
car & \textit{0.984 (0.006)}* & \textit{0.981 (0.008)} & \textit{\color{gray}0.82 (0.015)} & \textit{\color{gray}0.771 (0.028)} & \textit{\color{gray}0.951 (0.014)} & {0.976 (0.01)}\\
\hline
cardiotocography-3clases & {0.945 (0.012)} & \textit{0.949 (0.007)}* & \textit{\color{gray}0.896 (0.011)} & \textit{\color{gray}0.912 (0.012)} & \textit{\color{gray}0.913 (0.018)} & {0.942 (0.011)}\\
\hline
contrac & {0.545 (0.034)} & {0.56 (0.03)}* & \textit{\color{gray}0.525 (0.032)} & {0.559 (0.032)} & {0.553 (0.037)} & {0.557 (0.034)}\\
\hline
dermatology & \textit{\color{gray}0.969 (0.016)} & {0.975 (0.015)} & {0.972 (0.011)} & \textit{\color{gray}0.769 (0.024)} & {0.968 (0.019)} & {0.979 (0.014)}*\\
\hline
glass & {0.775 (0.073)}* & {0.742 (0.068)} & \textit{\color{gray}0.592 (0.087)} & \textit{\color{gray}0.642 (0.047)} & \textit{\color{gray}0.648 (0.085)} & {0.75 (0.035)}\\
\hline
heart-cleveland & {0.583 (0.042)} & {0.573 (0.045)} & {0.61 (0.04)} & {0.629 (0.04)}* & {0.624 (0.063)} & {0.608 (0.031)}\\
\hline
heart-va & \textit{0.372 (0.09)}* & {0.308 (0.068)} & {0.336 (0.087)} & {0.321 (0.046)} & {0.313 (0.088)} & {0.315 (0.091)}\\
\hline
iris & {0.96 (0.038)} & {0.96 (0.041)} & {0.96 (0.047)} & {0.95 (0.039)} & {0.963 (0.048)}* & {0.953 (0.053)}\\
\hline
lymphography & {0.871 (0.045)}* & {0.839 (0.048)} & {0.811 (0.048)} & {0.843 (0.061)} & \textit{\color{gray}0.754 (0.083)} & {0.836 (0.072)}\\
\hline
pittsburg-bridges-MATERIAL & {0.83 (0.086)} & {0.805 (0.08)} & {0.85 (0.058)} & {0.865 (0.047)}* & {0.82 (0.086)} & {0.865 (0.041)}*\\
\hline
pittsburg-bridges-REL-L & {0.705 (0.09)}* & {0.665 (0.106)} & {0.67 (0.086)} & {0.675 (0.059)} & {0.66 (0.084)} & {0.655 (0.08)}\\
\hline
pittsburg-bridges-SPAN & \textit{\color{gray}0.629 (0.1)} & \textit{\color{gray}0.594 (0.07)} & {0.659 (0.138)} & {0.694 (0.103)}* & {0.647 (0.088)} & {0.694 (0.117)}*\\
\hline
pittsburg-bridges-TYPE & \textit{0.674 (0.065)}* & {0.611 (0.09)} & {0.542 (0.108)} & {0.558 (0.067)} & {0.563 (0.086)} & {0.579 (0.05)}\\
\hline
seeds & {0.95 (0.035)} & {0.948 (0.022)} & {0.952 (0.019)} & {0.95 (0.018)} & {0.957 (0.033)}* & {0.95 (0.029)}\\
\hline
synthetic-control & {0.989 (0.009)} & \textit{\color{gray}0.972 (0.016)} & {0.987 (0.009)} & \textit{\color{gray}0.712 (0.022)} & {0.992 (0.008)}* & {0.985 (0.012)}\\
\hline
teaching & \textit{0.631 (0.098)}* & {0.583 (0.106)} & {0.531 (0.092)} & {0.545 (0.105)} & {0.514 (0.066)} & {0.541 (0.071)}\\
\hline
vertebral-column-3clases & {0.844 (0.032)} & {0.835 (0.031)} & {0.856 (0.039)}* & {0.821 (0.051)} & {0.848 (0.045)} & {0.85 (0.033)}\\
\hline
wine & {0.988 (0.021)}* & {0.982 (0.021)} & {0.979 (0.031)} & {0.971 (0.024)} & {0.974 (0.026)} & {0.982 (0.021)}\\
\hline
wine-quality-red & \textit{0.713 (0.025)}* & \textit{0.634 (0.015)} & {0.607 (0.018)} & \textit{\color{gray}0.581 (0.02)} & \textit{\color{gray}0.597 (0.015)} & {0.617 (0.022)}\\
\hline
\end{tabular}

}
\caption{Results of the classification study. Average out of sample accuracy is given along with its standard deviation in parantheses.  Asterisks denote the top performing method(s) for each dataset. Entries in italics have statistically significant difference in accuracy compared to cross-validated BART; those in gray have statistically significant differences and worse accuracy than BART.} 
\label{tab:classification}
\end{table}


Table ~\ref{tab:classification} reports the average and standard deviation of accuracy on the held-out validation datasets.  We tested the null hypothesis of no difference between BART-CV and each method using a paired Wilcoxon test. Table entries in italics were significantly different than BART-CV at $\alpha=0.05$, and the entries in gray had a statistically significant difference {\em and} worse estimated accuracy than BART.   Note that the paired design here lends this test some power even when the variability across random train/test splits is large relative to the estimated differences.  

It would be difficult to declare an overall ``winner'' from these results, even if we felt these 20 datasets were representative of a meaningful population of datasets. For example, of the 8 datasets in which the difference in out of sample accuracy between BART and random forests was statistically significant, random forests -- the closest competition -- had better accuracy in five and BART had better accuracy in three. As pointed out by a referee, if top classification accuracy is the goal we should probably ensemble methods or at least compare different classifiers on the particular dataset of interest. But we do note that in thirteen of the twenty datasets BART-CV was either a top performer or statistically indistinguishable from the best method. Of the remaining seven, only in two of these (car and wine-quality-red) was BART's performance worse than the second-best method with a statistically significant difference. If we did choose to ensemble methods, BART would be a natural candidate for that ensemble.

\begin{figure}[ht]
\begin{center}
{ \includegraphics[width=0.98\linewidth]{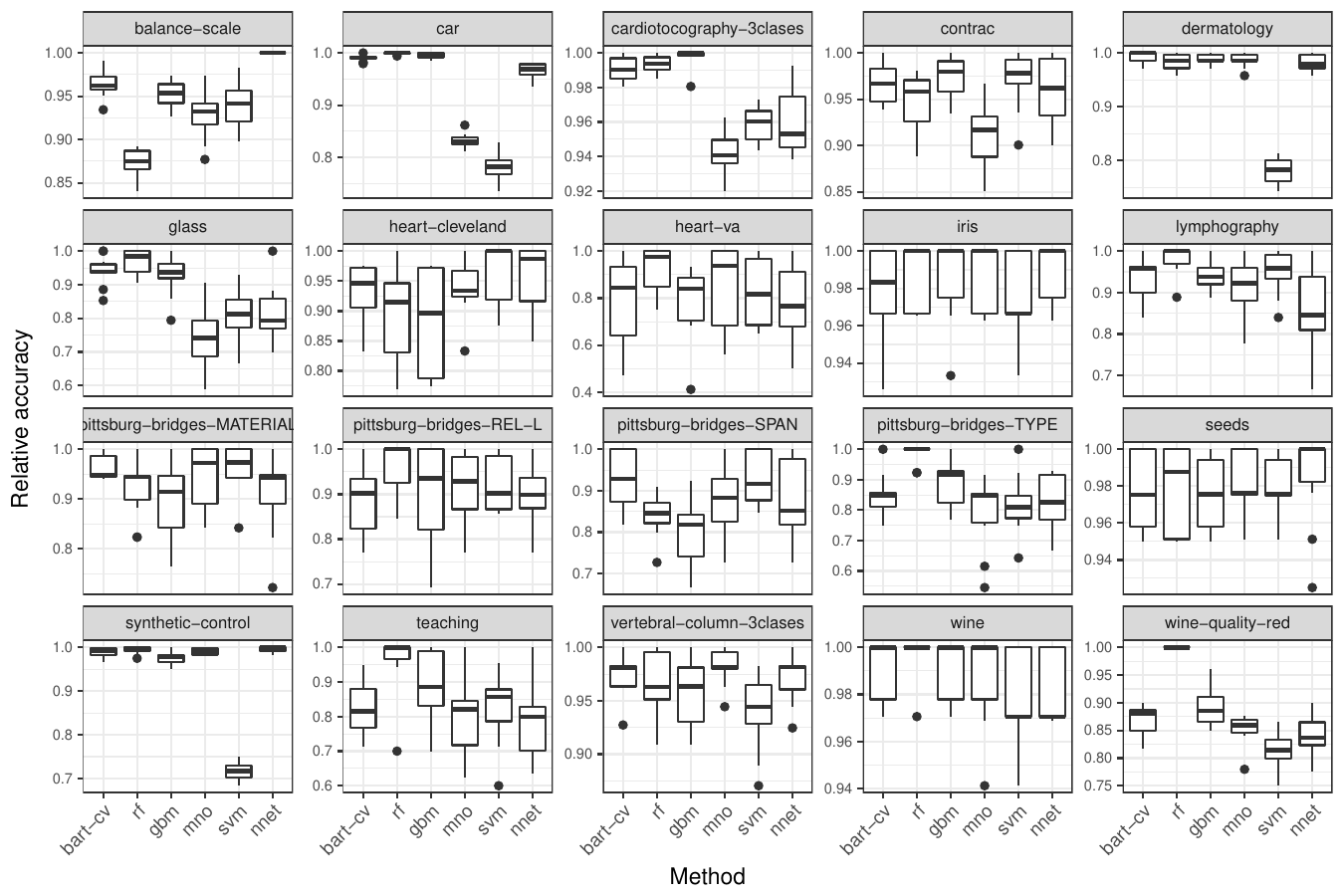}    
}
\end{center} 
\caption{Relative accuracy over the 10 splits for each dataset in the classification simulation.}
\label{fig:log-linear_relAcc}
\end{figure}

\begin{figure}[ht]
\begin{center}
{ \includegraphics[width=0.8\linewidth]{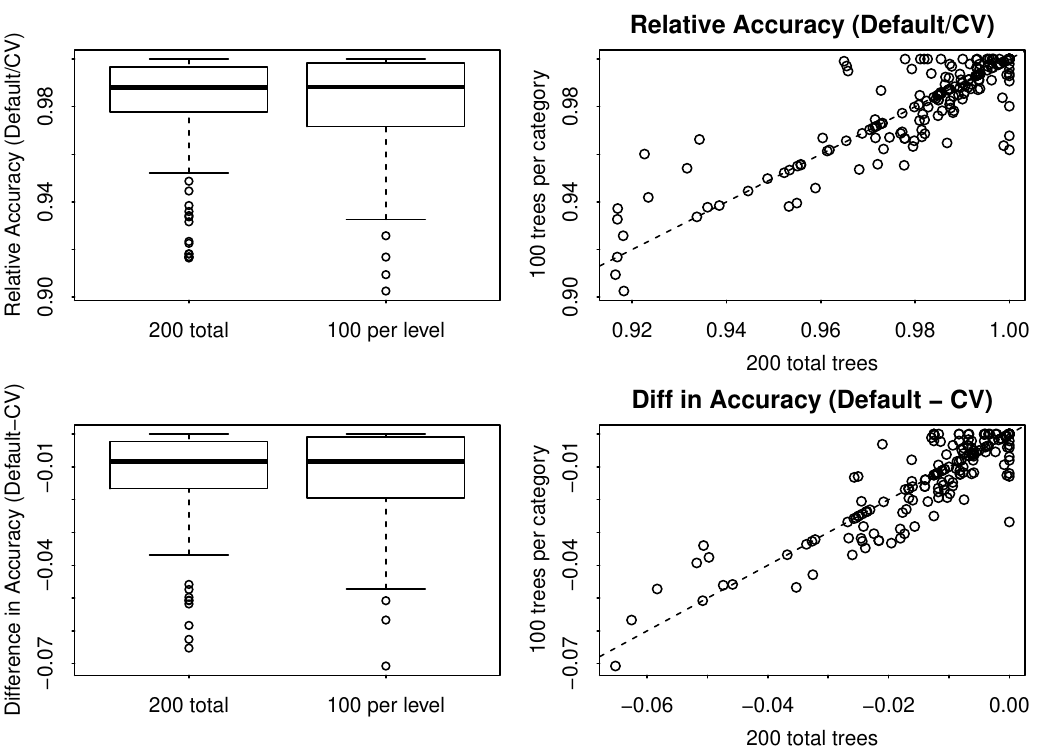}   
}
\end{center} 
\caption{Relative and absolute difference in accuracy of the two BART-default variants versus BART-CV across all folds of cross-validation.}
\label{fig:loglinear_defaults}
\end{figure} 

Figure~\ref{fig:log-linear_relAcc} gives a more nuanced view of these comparisons in light of the often substantial variability in out of sample accuracy across training and validation splits. It shows the accuracy of each method relative to the best performing method for each train/test split.  In most datasets the variability due to random train/test splits is larger than the gap between methods, save for a few datasets which consistently favor one method (e.g. neural nets in balance-scale and random forests in wine-quality-red).  We see here that for most datasets here BART is rarely far from the top-performing method for any given dataset train/test split.

Finally, Figure \ref{fig:loglinear_defaults} compares the relative accuracy of the two BART-default prior settings (200 total trees and 100 trees per outcome category) against BART-CV across folds of the cross-validation on the training dataset. This gives us $20\times10\times10= 2,000$ comparisons of the default BART settings versus the best parameter settings found over a grid search, albeit in slightly smaller datasets. Either default choices was nearly as accurate as the best parameter settings, within 2\% of the relative {\em or} absolute accuracy of the best model about 75\% of the time. There is no clear favorite between the two default models. The 100 trees per level setting was on average a little more accurate than using 200 total trees, but also more variable (as expected) and more computationally intensive to fit.

In summary, both cross-validated and default versions of BART have competitive predictive performance. Importantly for our purposes, the default variants are proper, fully Bayesian models that give valid posterior inference and may be incorporated into more complex models where cross-validation would be difficult or impossible. An immediate example of this is the binary logistic BART model embedded into the zero-inflated count regression model (where the partially latent binary variable $Z_i$ follows a distribution governed by a log-linear BART prior), which is explored in the next section.

\subsection{Example: Patent Citations}

When applying for new patents inventors must cite related existing patents, so the number of citations a patent receives is a (crude) measure of the invention's influence. We consider predicting citation counts using data from the European Patent Office (EPO) originally presented in \cite{Klein2015}. Several covariates are available; these are summarized in Table \ref{ta:patent-vars}. \cite{Klein2015} provide compelling evidence that these data cannot be adequately modeled without zero inflation and overdispersion, so we compare the ZINB-BART regression model to the semiparametric Bayesian ZINB regression models introduced in that paper.

\begin{table}
\centering
\caption{Summary of variables in the patent citation dataset.}
\label{ta:patent-vars}
\resizebox{\textwidth}{!}{%
\begin{tabular}{llllll}
Variable & \multicolumn{1}{c}{Description}                    & \multicolumn{1}{c}{Mean} & \multicolumn{1}{c}{SD} & \multicolumn{1}{c}{Min} & \multicolumn{1}{c}{Max} \\ \hline
opp      & Patent was opposed (1=yes, 0=no)                   & 0.41                     & -                            & 0                       & 1                       \\ \hline
biopharm & Patent from biopharmaceutical sector (1=yes, 0=no) & 0.44                     & -                            & 0                       & 1                       \\ \hline
ustwin   & U.S. ``twin'' patent exists (1=yes, 0=no)          & 0.61                     & -                            & 0                       & 1                       \\ \hline
patus    & Patent holder is from U.S (1=yes, 0=no)            & 0.33                     & -                            & 0                       & 1                       \\ \hline
patgsgr  & Patent holder is from Germany, Switzerland,        & 0.24                     & -                            & 0                       & 1                       \\
         & or Great Britain (1=yes, 0=no)                     &                          &                              &                         &                         \\ \hline
year     & Grant year                                         & -                        & -                            & 1980                    & 1997                    \\ \hline
ncountry & Number of designated states for the patent         & 7.8                      & 4.12                         & 1                       & 17                      \\ \hline
nclaims  & Number of claims against the patent                & 12.3                     & 8.13                         & 1                       & 50                      \\ \hline
ncit     & Number of citations of the patent                  & 1.6                      & 2.71                         & 0                       & 40                      \\ \hline
\end{tabular}
}
\end{table}

\cite{Klein2015} select a model based on stepwise selection using DIC under semiparametric regression models for the dispersion, zero-inflation, and mean parameters. Their selected model (\klein-1) is as follows:
\begin{align*}
\log[\f(x)] =&\ \beta_0^{\mu}+ \beta_1^{\mu}\text{opp} 
               + \beta_2^{\mu}\text{biopharm}
               + \beta_3^{\mu}\text{patus}
               + \beta_4^{\mu}\text{patsgr}\\
               &+ f_1^{\mu}(\text{ncountry})
               + f_2^{\mu}(\text{year})
               + f_3^{\mu}(\text{nclaims})\\
\logit[1-\omega(x)] =&\ \beta_0^{\omega}%
               + \beta_1^{\omega}\text{biopharm}
               + \beta_2^{\omega}\text{(year-1991)}
               + f_1^{\omega}(\text{ncountry})\\
\log[\kappa(x)] =&\ \beta_0^{\kappa} + \beta_1^{\kappa}\text{patus} + \beta_2^{\kappa}\text{patgsgr}.
\end{align*}
The functions $f_1^{\mu},f_2^{\mu},f_3^{\mu},$ and $f_1^{\omega}$ are modeled via cubic B-spline expansions using 20 knots, with shrinkage priors on the coefficients \citep{Klein2015}. We also consider two other specifications: A model that has the same specifications for $\f(\x)$ and $\omega(\x)$ as above but a constant $\kappa$ (\klein-2), and a ``saturated'' model that has a constant $\kappa$, and additive models for $\f(\x)$ and $\omega(\x)$ that include main effects for all categorical covariates and univariate B-spline basis expansions for each of the three continuous variables (\klein-3). We consider constant $\kappa$ models to compare results with ZINB-BART, which also uses a single dispersion parameter, and the ``saturated'' model is included to give some indication of the necessity of selection in this class of models. Prior distributions for the nonparametric components are the same as in \cite{Klein2015}. Posterior sampling was carried out via MCMC using the BayesX software package \citep{belitz2009bayesx}. 

As an alternative we consider a single ZINB-BART model with reasonable defaults - $\f(\x)$ has a log-linear BART prior with 200 trees and $a_0 = 2$, so that the marginal prior on $\mu(\x)$ puts approximately 95\% probability over the range $(0.02, 50)$ (specified by slightly inflating the $a_0$ that satisfies the heuristic in Section~\ref{sec:prior-count} with $y^*=50$). The excess zero probability $1-\omega(\x)$ has a  logistic BART prior with 200 total trees and $a_0=3.5\sqrt{2}$, so that $\Pr(|\logit[1-\omega(\x)]|<7)\approx 0.95$. The dispersion parameter $\kappa$ has a beta-prime prior with $a_\kappa=5,\ b_\kappa=3$, yielding a prior mode of 1, $E(\kappa)=2.5$, and $Var(\kappa) = 8.75$. The posterior mean of $\kappa$ was 1.16, with a 95\% credible interval of $(1.02,1.29)$, indicating strong support for overdispersion in the data.

\subsubsection{Results}

We apply the same outlier removal rule as \cite{Klein2015}, deleting observations with over 50 claims against them. (B-spline models are sensitive to outliers; ZINB-BART's tree-based basis functions are not and ZINB-BART's fits are essentially unchanged when including these points.) The models are evaluated based on two criteria: the Watanabe-Akaike/``widely applicable'' information criterion (WAIC) \citep{Watanabe2010-dd,Watanabe2013-fa}, defined as 

\beq
WAIC = -2 \left(\underbrace{\sum_{i=1}^n \log\left( \E[p(y_i\mid \x_i, \Theta)]\right)}_{LPD} - \underbrace{\sum_{i=1}^n  \Var[\log\{p(y_i\mid \x_i, \Theta)\}]}_{p_{waic}}\right),\label{eq:waic}
\eeq
where the expectations and variances are with respect to the posterior over $\Theta$ (overloading $\Theta$ for the moment to represent {\em all} the parameters, including any trees and their parameters, but marginalizing over any latent variables introduced for data augmentation). The first term inside the parentheses in \eqref{eq:waic} is the sum of the log predictive density at each data point (LPD), and the second term is a measure of the effective number of parameters ($p_{waic}$).
WAIC has a number of desirable features over other information criteria: As noted by \cite{Gelman2014-kj}, it averages over the posterior rather than conditioning on a point estimate, is invariant to reparameterization, and is more readily justified outside of regular parametric models. Under mild conditions model selection via WAIC is asymptotically equivalent to leave-one-out cross-validation. To more directly measure out of sample performance, we also estimated the log-loss (log-likelihood) of a single held out observation using ten fold cross-validation.

Table \ref{ta:waic} shows that ZINB-BART has the lowest WAIC and held-out log loss of all models considered, despite \klein-1 being chosen via stepwise selection and also being more flexible in some sense (by allowing the dispersion parameter $\kappa$ to vary with covariates). The estimated values of $p_{waic}$ show that all three \klein\ models have similar complexity, with the saturated model having approximately 11 additional effective parameters due to the additional nonlinear partial effects. However, this saturated model underperforms all the others -- the extra complexity swamps the mild increase in estimated predictive log likelihood. ZINB-BART has significantly more effective parameters (about 132 compared to 43-54) but a much higher predictive likelihood. The effective number of parameters is also far fewer than the actual number of parameters - a total of 400 regression trees and their associated leaf parameters, plus $\kappa$ -- due to the strong regularizing priors. 

\begin{table}
\centering

\begin{tabular}{l|l|l|l|l|}
\cline{2-5}
                                                                 & LPD    & $p_{waic}$ & WAIC              & CV-LL\\ \hline
\multicolumn{1}{|l|}{\klein-1 (stepwise DIC)}                    & -7783.5 & 43.6       & 15654.24          & -1.637 \\ \hline
\multicolumn{1}{|l|}{\klein-2 (stepwise DIC, constant $\kappa$)} & -7801.7 & 43.9       & 15691.14          & -1.640\\ \hline
\multicolumn{1}{|l|}{\klein-3 (saturated additive model, constant $\kappa$)}        & -7793.6 & 54.2       & 15695.48  & -1.640     \\ \hline
\multicolumn{1}{|l|}{ZINB-BART}                                  & -7688.2 & 131.5      & \textbf{15639.47} & \textbf{-1.631} \\ \hline
\end{tabular}
\caption{Comparison of the four models of the patent citation data. LPD is the in-sample log predictive density, marginalizing over any latent variables. $p_{waic}$ is the ``effective number of parameters'' in the definition of WAIC (defined in Eq~\eqref{eq:waic}). CV-LL is a ten-fold CV estimate of out of sample log loss for a single data point.}
\label{ta:waic}
\end{table}


%
%
%
%
%
%
%
%
%

%
%
%
%
%
%
%
%
%

\begin{figure}[ht]
\begin{center}
{ %
\includegraphics[width=0.8\linewidth]{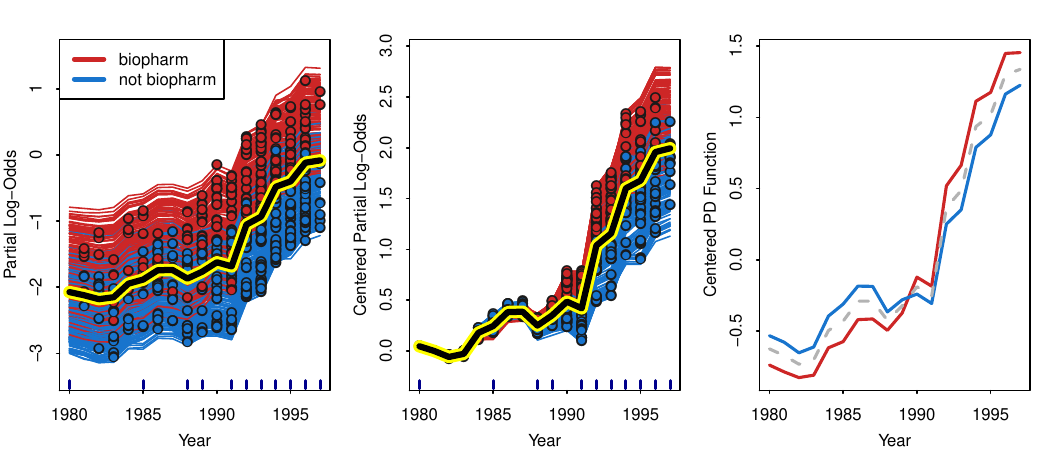} 
}
{ %
\includegraphics[width=0.8\linewidth]{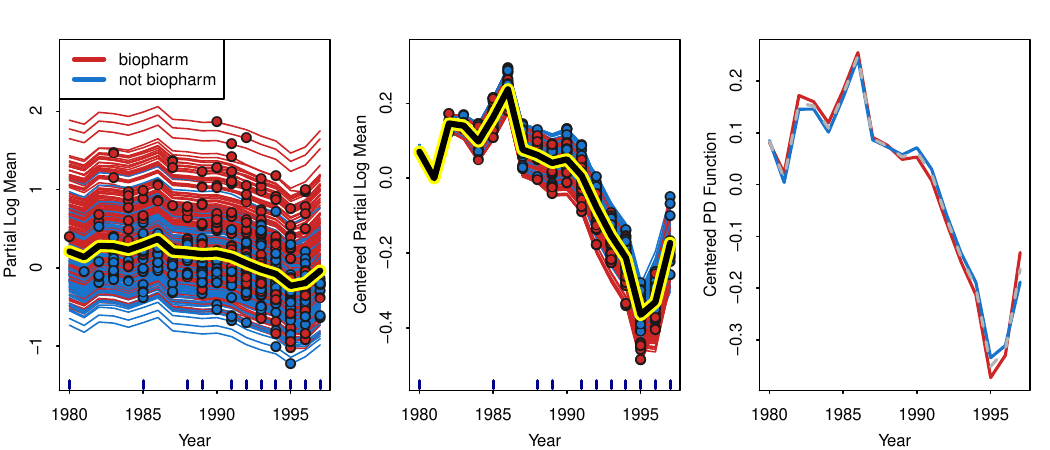}
}
\end{center}
\caption{Log-odds functions $\logit[1-\omega(\x)]$ are given in the top row, log mean functions $\log[\f(\x)]$ are in the bottom row. (Left) Partial dependence function (solid line) and a 10\% sample of the functions as year varies. (Center) The same as the right panel, except all curves are centered at their value in 1980. (Right) Partial dependence functions computed in the entire sample (dashed gray) and in biopharm/non-biopharm subgroups separately}
\label{fig:patent-logodds}
\end{figure}

Given an improvement in fit we might suspect that ZINB-BART is capturing some interactions that the additive models cannot. This does seem to be the case here. For example, there appears to be an interaction effect between biopharm and year in the excess zero process. This is supported by the existing literature; due to regulatory hurdles, biopharmaceutical innovations take more time to reach the market and be generally recognized \citep{jaffe1996flows}. Therefore we would expect to see a higher probability of an excess zero in recent years for biopharmaceutical patents, which is reflected in the ZINB-BART fit.

The first row of Figure~\ref{fig:patent-logodds} displays summaries of the posterior over $\logit[1-\omega(\x)]$, the log-odds of the conditional probability of an excess zero. In the leftmost plot the solid center line is the {\em partial dependence} (PD) function \citep{Friedman2001-vb} defined as
\beq
\hat{f_j}(t) = \frac{1}{n}\sum_{i=1}^n {\logit[1-\omega(\tilde \x_i)]},\nonumber
\eeq
where $\tilde x_{ik} = x_{ik}$ for $k\neq j$ and $x_{ij}=t$. Here the $j^{th}$covariate is year. As suggested by \cite{Goldstein2015-cj}, we also plot a 10\% sample of the individual response functions $f(\tilde\x_i)$, with dots indicating the actual year (PD plots alone can be misleading in the presence of interactions). The middle plot centers each of the curves at their 1980 value, which makes the interaction apparent: Recent biopharm patents are more likely to have excess zeros than non-biopharm patents. The rightmost plot displays mean-centered PD functions computed across the sample (in gray) and separately for biopharm/non-biopharm patents. The slope is much steeper for biopharm patents, indicating that the age of the patent is a more important factor for biopharm than non-biopharm patents. While we lack the means to do a formal hypothesis test, we can perform a crude ``placebo test'' by checking for this pattern in the mean of the negative binomial component, where it is not suggested by subject matter considerations. The second row of Figure~\ref{fig:patent-logodds} shows the same set of plots for $\log[\mu(\x_i)]$, where indeed no such pattern is apparent.

To conclude, we have seen that the ZINB-BART model fits better than additive semiparametric alternatives. This comes at some cost in summarizing and interpreting the fit, which would seem to be an advantage of the additive model. But of course the additive model could never capture the substantively interesting interaction we observed above without manual intervention. Further, fitting the additive model is not without its own challenges -- the results proposed by \cite{Klein2015} utilize stepwise selection on the {\em entire} dataset to select a model.  Subsequent inferences are not strictly valid from a Bayesian perspective due to the double use of the data, and we should not expect them to have frequentist validity either for the usual reasons that post-selection inference is invalid (see e.g. \cite{Berk2013-vn}).  Trying to search over potential interactions in addition to additive terms would compound this problem.  Fitting a single nonparametric model like ZINB-BART avoids this issue, and we have seen above that ZINB-BART can capture meaningful, interpretable interactions and nonlinearities that were not specified {\em a priori} without relying on explicit model selection or search. In this example the computational costs for each method are also similar; MCMC for ZINB BART took approximately 8 minutes on a recent MacBook, while fitting a single StAR model took about 4 minutes to obtain similar effective sample sizes for the linear predictors (not accounting for the stepwise model selection).

\section{Conclusion}\label{sec:conclusion}

We have introduced a novel prior and MCMC sampler that allow us to efficiently extend BART to log-linear models for unordered categorical and count responses. We expect that these models will be useful in a variety of settings, given the range of applied problems where the original BART model and its extensions have been successfully deployed. Like the original BART model, log-linear BART is highly modular and amenable to embedding within larger models for more complex applications. The use of a logistic regression BART model in the context of zero-inflated count data is just the one step in this direction.

These priors and algorithms can be used to fit a wide range of models including ordinal models like the continuation ratio logit, as well as hurdle versions of the Poisson and negative binomial models, with different data augmentation techniques. As another concrete example, in the supplemental material (Section~\ref{sec:hetmodel}) we describe how to fit models for continuous data with covariate-dependent heteroscedasticity using the methods in this paper. (This model using a slightly different prior distribution was presented by \cite{barthet}, concurrently with a presentation of methods in this paper and, is introduced in detail by \cite{Pratola2017-cz}.)

Another interesting extension is the use of shared trees across functions to borrow strength.  As noted by a referee, this could be particularly helpful in multinomial logistic regression with rare outcome categories.  One way to operationalize this is to use a common tree structure with distinct parameters for each function, which is essentially a BART prior with multivariate terminal node parameters. \cite{starling2018functional} and \cite{Linero2018-fw} take this approach in several different models. This requires only minor modification to the algorithms presented here (and none to the data augmentation) -- computation in such a shared-tree model with log-linear BART priors would proceed as in Algorithm~\ref{alg:backfit}, but using products of integrated likelihoods of the form in Eq.~\ref{eq:intlikexp}.  The downside of shared trees is that common trees need to be more complex than distinct trees if important covariates vary between the two functions, or if interactions between covariates differ across the regression functions (such as in the patent data).  With shared trees the strong shrinkage in the BART prior might prefer to smooth away these interesting features with modest sample sizes. Studying this tradeoff is an important area for future work.

Summarizing the fit of complicated nonparametric models like BART is also an important area for future work. Other authors -- beginning with CGM -- have proposed variable selection procedures for BART that could be applied in log-linear BART directly \citep{Bleich2014-lg}. Additionally, \cite{Linero} recently introduced a modification of CGM's tree prior that is more suitable for high-dimensional settings and provides a measure of variable importance.  This prior is immediately applicable to log-linear BART models, as is the interaction-detection variant of \cite{Du2018-jt}. Generic tools for providing interpretable summaries of complex nonparametric regression functions would be particularly useful.

Finally, there has recently been a surge of new results about the theoretical properties and frequentist operating characteristics of BART and Bayesian regression tree models in general \citep{Van_der_Pas2017-kr,Linero2017-ou,Rockova2017-qn,Rockova2018-tw}. These papers generally study the regression model with continuous outcomes and normal errors in Eq.~\eqref{eq:meanbart} under various modifications of CGM's tree prior. Extending these results to non-Gaussian models is a promising area for future research.

\bibliographystyle{./plainnat}
\bibliography{refs}

\end{document}


\def\spacingset#1{\renewcommand{\baselinestretch}%
{#1}\small\normalsize} \spacingset{1}

\if0\blind
{
  \title{\bf Supplemental Material for Log-Linear Bayesian Additive Regression Trees for Multinomial Logistic and Count Regression Models}
  \author{Jared S. Murray\thanks{
    The author gratefully acknowledges support from the National Science Foundation under grant numbers SES-1130706, SES-1631970 and DMS-1043903. Any opinions, findings, and conclusions or recommendations expressed
in this material are those of the author(s) and do not necessarily reflect the views of the funding agencies. Thanks to P. Richard Hahn and Carlos Carvalho for helpful comments and suggestions on an early version of this manuscript.
    }\hspace{.2cm}\\
    University of Texas at Austin\\
    }
  \maketitle
} \fi

\renewcommand\thesection{S.\arabic{section}}
\renewcommand\thefigure{S.\arabic{figure}}
\renewcommand\thetable{S.\arabic{figure}}

\if1\blind
{
  \bigskip
  \bigskip
  \bigskip
  \begin{center}
    {\LARGE\bf Supplemental Material for Log-Linear Bayesian Additive Regression Trees for Multinomial Logistic and Count Regression Models}
\end{center}
  \medskip
} \fi

\bigskip





\renewcommand{\theequation}{S.\arabic{equation}}

\section{Supplemental Material} 

\subsection{Parameterizing Logistic BART Models}\label{sec:lr-ex}

When $n_i=1$ for all $i$ and $c=2$, so that $y$ is a binary vector, we recover the binary regression model
\begin{equation}
p_{B}(y_i) =
\left(
\frac{
   \f^{(0)}(\x_i)%
}{
  \f^{(0)}(\x_i) + \f^{(1)}(\x_i)
}
\right)^{(1-y_i)}\left(
\frac{
   \f^{(1)}(\x_i)%
}{
  \f^{(0)}(\x_i) + \f^{(1)}(\x_i)%
}
\right)^{y_i},\label{eq:lr-ex-unid}
\end{equation}
which is a logistic regression model with log odds of success $\log[\f^{(1)}(\x_i)] - \log[\f^{(0)}(\x_i)]$. If $\f^{(0)}$ and $\f^{(1)}$ have the same number of trees (say $m$) and precision parameter $a_0^2$ then under our prior  $\tilde \f^{(1)}(\x_i) \eqd \f^{(1)}(\x_i)/\f^{(0)}(\x_i)$, where $\tilde \f^{(1)}$ has $2m$ trees and concentration parameter $2a_0^2$, so can write the model equivalently as 
\begin{equation}
p_{B}(y_i) =
\left(
\frac{
   1%
}{
  1 + \tilde\f^{(1)}(\x_i)
}
\right)^{(1-y_i)}\left(
\frac{
   \tilde\f^{(1)}(\x_i)%
}{
  1 + \tilde\f^{(1)}(\x_i)%
}
\right)^{y_i},\label{eq:lr-ex-id}
\end{equation}
in terms of the identified parameter $\tilde \f^{(1)}(\x_i)$. The prior and likelihood (and therefore the posterior) are identical, but the performance of the data augmented MCMC algorithm can be substantially different for extreme probabilities. 

To illustrate we consider a simple synthetic example. The probabilities are given by

\beq
\Pr(y_i\mid \x_i=x) = \exp[\f^*(\x_i)]/(1+\exp[\f^*(\x_i)]),\quad \f^*(x) = 12(\x_i-0.5)\nonumber
\eeq
so that the true log odds range over $\pm 6$, yielding probabilities in $(0.0025, 0.9975)$. The covariates are placed (not sampled) uniformly over $(0,1)$. 
In addition to the identified and unidentified logit models we compare the BART probit model introduced by CGM, which assumes that
\beq
\Pr(y_i=1\mid \x_i) = \Phi[\f_{CGM}(\x_i)]\nonumber
\eeq
where $\f_{CGM}(\cdot)$ has the original BART prior with leaf parameters $\mu_{ht}\sim N(0, 1.5^2/m)$, so that $\Pr(|\f(\x_i)|<3) \equiv \Pr(\Phi[-3]<\Phi[\f(\x_i)]<\Phi[3])=0.95$ \emph{a priori}. This is approximately the true range of the probabilities in our synthetic example, and we use the same condition to set $a_0$ in the logistic models. 

We generated 25 datasets of size $n=100$ and ran the MCMC algorithm for 6,000 iterations, discarding the first 1,000 as burn-in. We estimate the log odds function at each covariate value. They are given by $\log[\f^{(1)}(\x_i)] - \log[\f^{(0)}(\x_i)]$ for the unidentified logit model, $\log[\tilde\f^{(1)}(\x_i)]$ for the identified logit model, and  $\log(\Phi[\f(\x_i)]) - \log(1-\Phi[\f(\x_i)])$ for BART probit.
We compare the average effective sample size of the log-odds function over the 25 replicates.  The effective sample size is given by 
\[
\frac{T}{1+2\sum_{k=1}^\infty \rho_k},
\]
where $T$ is the length of the MCMC chain and $\rho_k$ is the lag $k$ autocorrelation. It is the number of independent samples required to reach the same level of sampling error as the correlated MCMC draws. This is estimated using the R package \texttt{coda} \citealt{coda-r}. Unlike the two logit models, BART probit has a different target distribution and therefore the effective sample sizes are not directly comparable. We include it in the comparison primarily to illustrate the operating characteristics of a similar, well-known data augmentation scheme.

\begin{figure}[ht]
\begin{center}
{ \includegraphics[width=0.6\linewidth]{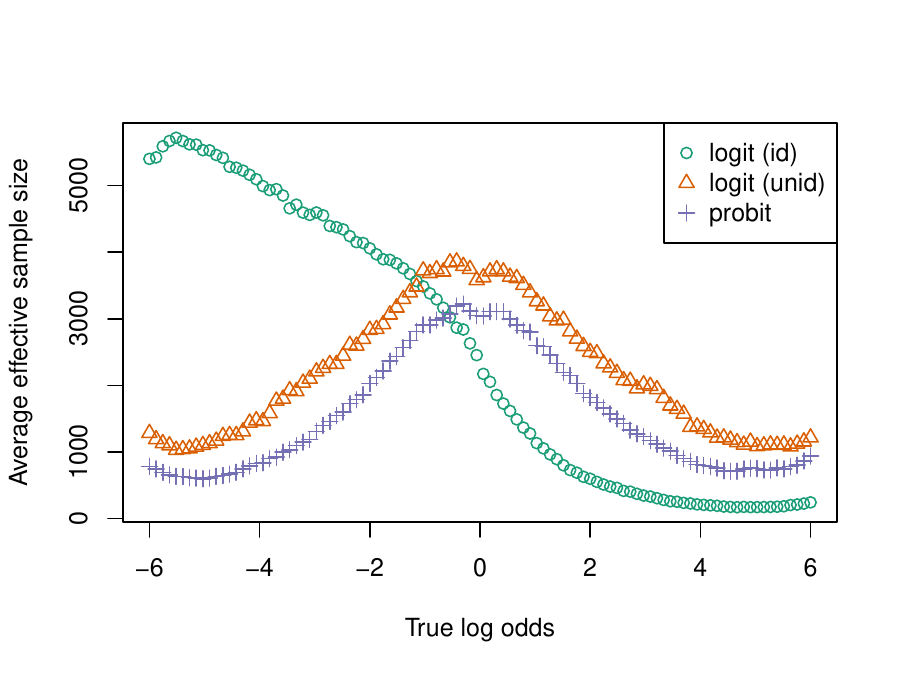}
}
\end{center}
\caption{Average effective sample size for estimates of the log odds for the simulated datasets in Section \ref{sec:lr-ex}. The labels ``logit (id)'' and ``logit (unid)'' refer to the identified/unidentified parameterizations of the logistic regression model (Eqs \eqref{eq:lr-ex-id} and \eqref{eq:lr-ex-unid}, respectively).}
\label{fig:lr-ess}
\end{figure} 

Figure \ref{fig:lr-ess} shows the results. The two logit models perform much differently; the identified model mixes extremely well when the log odds are small and extremely poorly when they are large. The unidentified model mixes best near $0$, degrading as the log odds increase in magnitude. But the unidentified parameterization has a minimum average ESS of about 1,030, compared to 170 for the identified parameterization, so the unidentified paramterization has the benefit of performing adequately everywhere.

Examining the two data augmentation schemes sheds some light on this behavior. In the identified parameterization, the latent variables are sampled $\phi_i\sim Gamma(1, 1+\tilde\f^{(1)}(\x))$ prior to updating $\tilde\f^{(1)}$. When $\tilde\f^{(1)}(\x_i)$ is large, the variance of this full conditional is small and $\phi_i$ will make small moves leading to high cross-correlation between the parameters determining $\f^{(1)}(\x_i)$ and $\phi_i$. When $\tilde\f^{(1)}(\x_i)$ is small the full conditional is much more dispersed, reducing the crosscorrelation and leading to the excellent behavior in Fig \ref{fig:lr-ess}. In the unidentified parameterization, $\phi_i\sim Gamma(1, \f^{(0)}(\x_i)+\f^{(1)}(\x))$. For probabilities near 1, increases in $\f^{(1)}(\x_i)$ are offset by compensatory decreases in $\f^{(0)}(\x_i)$ (which is fixed at 1 in the identified sampler), since our prior for both functions is centered at 1. However, for probabilities at either extreme $|\f^{(0)}(\x_i)+\f^{(1)}(\x)|$ will tend to be large and induce a greater degree of crosscorrelation. For probabilities near 0.5, $f^{(0)}(\x_i)\approx f^{(1)}(\x_i)$ and fit is freely allocated between $\f^{(1)}$ and $\f^{(0)}$. 
Hence MCMC in the unidentified parameterization behaves similarly to the BART-probit sampler, which is constructed using \cite{AlbertChib}'s data augmentation:
\begin{enumerate}
\item Sample $\phi_i\sim N(\f(\x_i), 1)\ind{\phi_i>0}$ if $y_i=1$ or $\phi_i\sim (\f(\x_i), 1)\ind{\phi_i<0}$ if $y_i=0$
\item Update $\f(\cdot)$ via the CGM MCMC algorithm
\end{enumerate}

Here the crosscorrleation between the parameters determining $\f(\x_i)$ and $\phi_i$ is weakest when $\f(\x_i)\approx 0$, or $0$ on the log odds scale.

Our results suggest that the identified parmaterization of the logit model mixes more efficiently for log odds less than about $-1$ (probabilities less than about $0.27$). Therefore the most efficient parameterization will depend on factors including the balance of the outcome as well as the distribution of the covariates and their discriminative power. 
If it is known that the outcome is rare and the predictors relatively weak working in the identified parameterization may be more efficient. In the absence of such strong prior knowledge the unidentified parameterization yields good results across a range of settings, and is more sensible as a default. In the case of the multinomial regression model it also avoids the risk of accidentally specifying a poor prior through an inappropriate choice of reference category, since all outcome values are treated symmetrically.

\subsection{Proof of propositions}
\label{sec:app-pars}
\label{sec:app-zinb-aug}

\textbf{Proposition~\ref{prop:zinbaug}:}
Collecting terms in the augmented variables, we have:

\begin{align*}
p_{ZINB}(y_i, Z_i, \phi_i, \xi_i &\mid \f^{(0)}, \f_i, \kappa, \f)\\
=& 
\f^{(0)}(\x_i)^{1-Z_i}\f^{(1)}(\x_i)^{Z_i}\exp[-\phi_i\{\f^{(0)}(\x_i) + \f^{(1)}(\x_i)\}]\\
&\left\{
\frac{1}{\Gamma(\rr)y_i!}
   \rr^\rr%
   [\mu_{0i}\f(\x_i)]^{y_i}%
\xi_i^{\rr+y_i-1}\exp\left[-\xi_i( \rr + \mu_{0i}\f(\x_i))\right]
\right\}^{Z_i}\\
&\ind{Z_i=1\text{ if }y_i>0}
.
\end{align*}

Since $\int_0^\infty t^{u-1}\exp(-st)dt = \Gamma(u)/s^u$,
\begin{align*}
\int_0^\infty\int_0^\infty p_{ZINB}&(y_i, Z_i, \phi_i, \xi_i \mid \f^{(0)}, \f_i, \kappa, \f)d\phi_i d\xi_i\\
=& 
\f^{(0)}(\x_i)^{1-Z_i}\f^{(1)}(\x_i)^{Z_i}\int_0^\infty\exp[-\phi_i\{\f^{(0)}(\x_i) + \f^{(1)}(\x_i)\}]d\phi_i\\
&\times\left\{
\frac{1}{\Gamma(\rr)y_i!}
   \rr^\rr%
   [\mu_{0i}\f(\x_i)]^{y_i}%
\int_0^\infty\xi_i^{\rr+y_i-1}\exp\left[-\xi_i( \rr + \mu_{0i}\f(\x_i))\right]d\xi_i
\right\}^{Z_i}\\
&\times\ind{Z_i=1\text{ if }y_i>0}\\
=& 
\frac{\f^{(0)}(\x_i)^{1-Z_i}\f^{(1)}(\x_i)^{Z_i}}{\f^{(0)}(\x_i) + \f^{(1)}(\x_i)}\\
&\times\left\{
\frac{1}{\Gamma(\rr)y_i!}
   \rr^\rr%
   [\mu_{0i}\f(\x_i)]^{y_i}%
\frac{\Gamma(\rr+y_i)}{( \rr + \mu_{0i}\f(\x_i))^{\rr+y_i}}
\right\}^{Z_i}\\
&\times\ind{Z_i=1\text{ if }y_i>0}\\
=& 
\frac{\f^{(0)}(\x_i)^{1-Z_i}\f^{(1)}(\x_i)^{Z_i}}{\f^{(0)}(\x_i) + \f^{(1)}(\x_i)}\\
&\times\left\{
\frac{\Gamma(\rr+y_i)}{\Gamma(\rr)y_i!}
   \left(\frac{\rr}{\rr + \mu_{0i}\f(\x_i)}\right)^\rr%
   \left(\frac{\mu_{0i}\f(\x_i)}{\rr + \mu_{0i}\f(\x_i)}\right)^{y_i}
\right\}^{Z_i}\\
&\times\ind{Z_i=1\text{ if }y_i>0}
.\label{eq:zipaug02}
\end{align*}
To sum over $Z_i$, consider the two cases $y_i>0$ and $y_i=0$. If $y_i=0$ then
\begin{align*}
\sum_{Z_i=0}^1\int_0^\infty\int_0^\infty p_{ZINB}&(y_i, Z_i, \phi_i, \xi_i \mid \f^{(0)}, \f_i, \kappa, \f)d\phi_i d\xi_i \\
=& 
\frac{\f^{(1)}(\x_i)}{\f^{(0)}(\x_i) + \f^{(1)}(\x_i)}
\frac{\Gamma(\rr+y_i)}{\Gamma(\rr)y_i!}
   \left(\frac{\rr}{\rr + \mu_{0i}\f(\x_i)}\right)^\rr%
   \left(\frac{\mu_{0i}\f(\x_i)}{\rr + \mu_{0i}\f(\x_i)}\right)^{y_i}
\end{align*}
due to the indicator function. Otherwise if $y_i=1$ then 
\begin{align*}
\sum_{Z_i=0}^1\int_0^\infty\int_0^\infty p_{ZINB}&(y_i, Z_i, \phi_i, \xi_i \mid \f^{(0)}, \f_i, \kappa, \f)d\phi_i d\xi_i \\
=& \frac{\f^{(0)}(\x_i)}{\f^{(0)}(\x_i) + \f^{(1)}(\x_i)}\\
&+\frac{\f^{(1)}(\x_i)}{\f^{(0)}(\x_i) + \f^{(1)}(\x_i)}
\frac{\Gamma(\rr+y_i)}{\Gamma(\rr)y_i!}
   \left(\frac{\rr}{\rr + \mu_{0i}\f(\x_i)}\right)^\rr%
   \left(\frac{\mu_{0i}\f(\x_i)}{\rr + \mu_{0i}\f(\x_i)}\right)^{y_i}
\end{align*}
So we have 
\begin{align*}
\sum_{Z_i=0}^1\int_0^\infty\int_0^\infty p_{ZINB}&(y_i, Z_i, \phi_i, \xi_i \mid \f^{(0)}, \f_i, \kappa, \f)d\phi_i d\xi_i \\
=& \frac{\f^{(0)}(\x_i)}{\f^{(0)}(\x_i) + \f^{(1)}(\x_i)}\ind{y_i=0}\\
&+\frac{\f^{(1)}(\x_i)}{\f^{(0)}(\x_i) + \f^{(1)}(\x_i)}
\frac{\Gamma(\rr+y_i)}{\Gamma(\rr)y_i!}
   \left(\frac{\rr}{\rr + \mu_{0i}\f(\x_i)}\right)^\rr%
   \left(\frac{\mu_{0i}\f(\x_i)}{\rr + \mu_{0i}\f(\x_i)}\right)^{y_i}
\end{align*}
as required.

\noindent\textbf{Proposition~\ref{prop:pars}:} To set the parameters $\pc$ and $\pd$ from $a_0$ and $m$, note that
\begin{align*}
\E[\log(\lambda_{ht})] &= 0 \\
\Var[\log(\lambda_{ht})] &= \loggamma''(\pc) + [\loggamma'(\pc) - \log(\pd) ]^2,
\end{align*}
where $\psi(\pc) = \log[\Gamma(\pc)]$.
Enforcing $\Var[\log(\lambda_{ht})]=a^2_0/m$ requires that $\pc,\ \pd$ solve
\beq
\loggamma''(\pc) + [\loggamma'(\pc) - \log(\pd) ]^2 - a^2_0/m = 0.\label{eq:mixquad}\nonumber
\eeq
The real roots of \eqref{eq:mixquad} are given by 
\begin{equation*}
\pd = \exp\left[ 
  \sqrt{a^2_0/m - \loggamma''(\pc)}  \pm \loggamma'(\pc)
\right],
\end{equation*}
subject to $a^2_0/m - \loggamma''(\pc)\geq 0$. Taking $a^2_0/m - \loggamma''(\pc)= 0$ minimizes $\pd$, which concentrates more mass around zero on the log scale and is an appropriate choice for a strong regularizing prior. So $\pc$ is obtained numerically as the solution to $\loggamma''(\pc) = a_0^2/m$, which is trivial as $\loggamma''(\pc)$ is monotonically decreasing, and $\pd = \exp[\loggamma'(\pc)]$. 

\noindent\textbf{Proposition~\ref{prop:parsapprox}:} The exact solutions for the parameters in Proposition~\ref{prop:pars} can be approximated by $\pc\approx m/a^2_0 + 0.5$ and $\pd\approx m/a^2_0$.
Let $v = a^2_0/m$. Typically $v$ will be quite small, so $\pc$ will be large. The Laurent series of $\loggamma''(\pc)$ at $\pc=\infty$ is 
\beq
\frac{1}{\pc} + \frac{1}{2\pc^2} + O\left(\frac{1}{\pc^3}\right)\nonumber
\eeq
Using the first two terms of the series to approximate $\loggamma''(\pc)$ we want to solve
$v = \frac{1}{\pc} + \frac{1}{2\pc^2}$. Since $v,c$ are both positive, the only solution is
\beq
\pc = \frac{1+\sqrt{1+2v}}{2v}\nonumber
\eeq
We can obtain a simpler expression with one more approximation:
\begin{align*}
\pc &= \frac{1+\sqrt{1+2v}}{2v}\\
&= \frac{1}{2v} + \sqrt{\frac{1}{4v^2} + \frac{1}{2v}}\\
&\approx \frac{1}{2v} + \sqrt{\frac{1}{4v^2} + \frac{1}{2v} + \frac{1}{4}}%
\\
&= \frac{1}{2v} + \sqrt{\left(\frac{1}{2v} + \frac{1}{2}\right)^2}\\
&= \frac{1}{v} + \frac{1}{2} \\
&= \frac{m}{a_0^2} + \frac{1}{2}
\end{align*}
The expansion of $\exp[\loggamma'(\pc)]$ at $\pc=\infty$ is $\pc - 0.5 + O(1/\pc)$, so $\pd\approx m/a^2_0$. Thus when $m>> a_0^2$, we have $\pc\approx m/a_0^2+0.5$ and $\pd\approx m/a_0^2$. For all the settings of $m$ and $a_0$ considered in this paper, the largest relative error under these approximations is less than 2\% for both $\pc$ and $\pd$. These include some extreme settings from the cross validation exercise, however, and the approximation is usually much better. For example, the multinomial logistic regression default parameter setting $m=100, a_0 = 3.5/\sqrt{2}$ yields an approximation with less then $0.03\%$ relative error.

\subsection{MCMC for ZINB-BART}\label{sec:zinb-mcmc}

A single step of the ZINB MCMC algorithm proceeds as follows: 
\begin{enumerate}

\item Block update $(\rr, Z, \xi, \phi\mid -)$ by composition. (These steps are order-dependent.)
\begin{enumerate}
\item First sample $\kappa$ from
\beq
p(\rr\mid y, \x, \f, f^{(0)}, f^{(1)}, \rr)\propto p(\rr)
\prod_{i=1}^n
p_{ZINB}(y_i\mid \x_i, \f, f^{(0)}, f^{(1)}, \rr)\nonumber
\eeq
using a Metropolis-Hastings step (we use a random walk on the log scale, with a Gaussian proposal tuned to give acceptance rates of about 0.23).

\item Given the new value for $\rr$, sample $Z$. The $Z_i$'s are mutually independent given $\rr, \omega$ and $\f$. If $y_i>0$, $Z_i=1$. %
Otherwise $p(Z_i\mid \rr, \omega, \f)$ is Bernoulli with probability
\beq
\frac{
  \omega(\x_i)\pnb(0\mid \x_i, \rr, \f)
}{
  1-\omega(\x_i) + \omega(\x_i)\pnb(0\mid \x_i, \rr, \f)
}.\nonumber
\eeq
\item Finally $(\xi, \phi)$ are sampled from their joint full conditional. This is particularly simple due to their conditional independence: For all observations with $Z_i=1$, sample $\xi_i$ independently from
\beq
(\xi_i\mid -) \sim G(\kappa+y_i, \rr+\mu_{0i}\f(\x_i)),\nonumber
\eeq
and for each $1\leq i\leq n$ sample
\beq
(\phi_i\mid -) \sim Exp(\f^{(0)}(\x_i) + \f^{(1)}(\x_i)).\nonumber
\eeq
\end{enumerate}

\item Update $f^{(j)}$ for $j\in \{0,1\}$ using Algorithm \ref{alg:backfit} and the expressions in Section \ref{sec:mcmc} with
\begin{equation*}
r_{ht} = \sum_{i: \x_i\in A^{(j)}_{ht}} \ind{Z_i = j},\quad 
s_{ht} = \sum_{i: \x_i\in A^{(j)}_{ht}} \phi_{i} \noth^{(j)}(\x_i)
\end{equation*}
where $\noth^{(j)}(\x_i) = \prod_{l\neq h} g(\x, T^{(j)}_h, \Lambda^{(j)}_h)$ is the fit from all but the $h^{th}$ tree.

\item Update $\f$ using Algorithm \ref{alg:backfit} and the expressions in Section \ref{sec:mcmc} with
\begin{equation*}
r_{ht} = \sum_{i: \x_i\in A_{ht}} Z_iy_{i},\quad 
s_{ht} = \sum_{i: \x_i\in A_{ht}} Z_i\xi_{i}\mu_{0i} \noth(\x_i)
\end{equation*}
where $\noth(\x_i) =\prod_{l\neq h} g(\x, T_h, \Lambda_h)$ is the fit from all but the $h^{th}$ tree.
\end{enumerate}
Note that all three regression functions can be updated in parallel, as they are conditionally independent given the latent variables.

\subsection{Additional Classification Study Results}

{ \scriptsize
\begin{longtable}{l|r|r|r|r|r|r|r|r|r|r|l|}
\cline{2-12}
                                                 & \multicolumn{1}{c|}{rf} & \multicolumn{4}{c|}{gbm}             & \multicolumn{1}{c|}{mno} & \multicolumn{1}{c|}{svm} & \multicolumn{2}{c|}{nnet} & \multicolumn{2}{c|}{bart-cv} \\ \cline{2-12} 
                                                 & mtry                    & n.trees & int depth & shrinkage & n  & decay                    & C                        & size        & decay       & ntree     & $a_0\sqrt{2}$    \\ \hline
\hline
balance-scale & 2 & 150 & 1 & 0.1 & 10 & 1e-04 & 1.00 & 5 & 1e-01 & 100 & 6.0\\
\hline
 & 2 & 50 & 2 & 0.1 & 10 & 1e-04 & 0.50 & 5 & 1e-01 & 25 & 6.0\\
\hline
 & 2 & 100 & 2 & 0.1 & 10 & 1e-04 & 1.00 & 5 & 0e+00 & 25 & 6.0\\
\hline
 & 2 & 100 & 2 & 0.1 & 10 & 1e-04 & 1.00 & 5 & 1e-01 & 25 & 6.0\\
\hline
 & 2 & 100 & 1 & 0.1 & 10 & 1e-04 & 0.25 & 5 & 1e-01 & 66 & 6.0\\
\hline
 & 2 & 150 & 1 & 0.1 & 10 & 1e-04 & 0.25 & 5 & 1e-01 & 25 & 6.0\\
\hline
 & 2 & 150 & 2 & 0.1 & 10 & 1e-04 & 1.00 & 5 & 1e-01 & 66 & 6.0\\
\hline
 & 2 & 150 & 1 & 0.1 & 10 & 1e-04 & 0.25 & 5 & 1e-01 & 25 & 6.0\\
\hline
 & 2 & 150 & 2 & 0.1 & 10 & 1e-04 & 1.00 & 5 & 1e-01 & 25 & 6.0\\
\hline
 & 2 & 150 & 2 & 0.1 & 10 & 1e-04 & 0.25 & 5 & 1e-01 & 25 & 6.0\\
\hline
car & 6 & 150 & 3 & 0.1 & 10 & 1e-04 & 1.00 & 5 & 1e-01 & 25 & 6.0\\
\hline
 & 6 & 100 & 3 & 0.1 & 10 & 1e-01 & 1.00 & 5 & 0e+00 & 25 & 6.0\\
\hline
 & 6 & 150 & 3 & 0.1 & 10 & 1e-04 & 1.00 & 5 & 1e-04 & 25 & 6.0\\
\hline
 & 6 & 150 & 3 & 0.1 & 10 & 1e-04 & 1.00 & 5 & 1e-01 & 25 & 6.0\\
\hline
 & 6 & 150 & 3 & 0.1 & 10 & 0e+00 & 1.00 & 5 & 1e-01 & 25 & 6.0\\
\hline
 & 6 & 150 & 3 & 0.1 & 10 & 1e-04 & 1.00 & 5 & 1e-01 & 25 & 6.0\\
\hline
 & 6 & 150 & 3 & 0.1 & 10 & 1e-04 & 1.00 & 5 & 1e-01 & 50 & 6.0\\
\hline
 & 6 & 150 & 3 & 0.1 & 10 & 1e-04 & 1.00 & 5 & 1e-04 & 25 & 6.0\\
\hline
 & 6 & 150 & 3 & 0.1 & 10 & 1e-04 & 1.00 & 5 & 1e-01 & 25 & 6.0\\
\hline
 & 6 & 150 & 3 & 0.1 & 10 & 1e-04 & 0.50 & 5 & 1e-01 & 25 & 6.0\\
\hline
cardiotocography-3clases & 11 & 150 & 3 & 0.1 & 10 & 1e-01 & 1.00 & 5 & 1e-01 & 66 & 6.0\\
\hline
 & 11 & 150 & 2 & 0.1 & 10 & 1e-01 & 1.00 & 3 & 1e-01 & 66 & 6.0\\
\hline
 & 11 & 150 & 3 & 0.1 & 10 & 1e-01 & 1.00 & 5 & 1e-01 & 66 & 6.0\\
\hline
 & 11 & 150 & 2 & 0.1 & 10 & 1e-01 & 1.00 & 5 & 1e-01 & 100 & 6.0\\
\hline
 & 11 & 150 & 3 & 0.1 & 10 & 1e-01 & 1.00 & 3 & 1e-01 & 100 & 6.0\\
\hline
 & 11 & 150 & 3 & 0.1 & 10 & 1e-04 & 1.00 & 5 & 1e-01 & 25 & 6.0\\
\hline
 & 11 & 150 & 3 & 0.1 & 10 & 1e-01 & 1.00 & 5 & 1e-01 & 100 & 6.0\\
\hline
 & 11 & 150 & 3 & 0.1 & 10 & 1e-04 & 1.00 & 5 & 1e-01 & 66 & 6.0\\
\hline
 & 11 & 150 & 3 & 0.1 & 10 & 1e-01 & 1.00 & 5 & 1e-01 & 100 & 6.0\\
\hline
 & 11 & 150 & 3 & 0.1 & 10 & 1e-01 & 1.00 & 5 & 1e-01 & 66 & 6.0\\
\hline
contrac & 2 & 100 & 2 & 0.1 & 10 & 1e-01 & 1.00 & 3 & 1e-01 & 66 & 3.5\\
\hline
 & 2 & 50 & 3 & 0.1 & 10 & 1e-01 & 0.50 & 3 & 1e-01 & 25 & 2.0\\
\hline
 & 2 & 50 & 3 & 0.1 & 10 & 1e-01 & 1.00 & 5 & 1e-04 & 25 & 6.0\\
\hline
 & 2 & 100 & 2 & 0.1 & 10 & 1e-01 & 1.00 & 5 & 1e-01 & 66 & 2.0\\
\hline
 & 2 & 100 & 1 & 0.1 & 10 & 1e-01 & 0.50 & 5 & 1e-01 & 25 & 2.0\\
\hline
 & 2 & 50 & 2 & 0.1 & 10 & 1e-04 & 0.50 & 3 & 1e-01 & 66 & 6.0\\
\hline
 & 2 & 50 & 2 & 0.1 & 10 & 1e-01 & 1.00 & 3 & 1e-01 & 25 & 6.0\\
\hline
 & 2 & 50 & 1 & 0.1 & 10 & 1e-01 & 1.00 & 5 & 1e-01 & 25 & 2.0\\
\hline
 & 2 & 100 & 3 & 0.1 & 10 & 1e-01 & 0.50 & 3 & 1e-01 & 100 & 6.0\\
\hline
 & 2 & 100 & 2 & 0.1 & 10 & 1e-04 & 0.50 & 3 & 1e-01 & 25 & 3.5\\
\hline
dermatology & 2 & 50 & 3 & 0.1 & 10 & 1e-01 & 0.25 & 5 & 1e-01 & 25 & 6.0\\
\hline
 & 2 & 50 & 3 & 0.1 & 10 & 1e-01 & 0.25 & 5 & 1e-01 & 25 & 3.5\\
\hline
 & 2 & 50 & 3 & 0.1 & 10 & 1e-01 & 0.25 & 3 & 1e-01 & 33 & 6.0\\
\hline
 & 34 & 100 & 2 & 0.1 & 10 & 0e+00 & 0.25 & 5 & 1e-04 & 33 & 3.5\\
\hline
 & 2 & 50 & 3 & 0.1 & 10 & 1e-01 & 0.25 & 5 & 1e-01 & 100 & 2.0\\
\hline
 & 2 & 50 & 1 & 0.1 & 10 & 0e+00 & 0.25 & 5 & 1e-04 & 25 & 6.0\\
\hline
 & 18 & 50 & 2 & 0.1 & 10 & 0e+00 & 0.25 & 5 & 1e-01 & 25 & 3.5\\
\hline
 & 2 & 100 & 1 & 0.1 & 10 & 1e-01 & 0.25 & 5 & 1e-01 & 100 & 2.0\\
\hline
 & 2 & 50 & 3 & 0.1 & 10 & 1e-01 & 0.25 & 3 & 1e-01 & 100 & 6.0\\
\hline
 & 2 & 50 & 2 & 0.1 & 10 & 0e+00 & 0.25 & 5 & 1e-01 & 100 & 2.0\\
\hline
glass & 2 & 100 & 3 & 0.1 & 10 & 1e-04 & 0.25 & 5 & 1e-01 & 25 & 6.0\\
\hline
 & 2 & 100 & 3 & 0.1 & 10 & 0e+00 & 0.50 & 5 & 1e-01 & 33 & 6.0\\
\hline
 & 5 & 150 & 3 & 0.1 & 10 & 1e-04 & 0.25 & 5 & 1e-01 & 25 & 6.0\\
\hline
 & 2 & 150 & 2 & 0.1 & 10 & 1e-01 & 0.50 & 5 & 1e-01 & 100 & 6.0\\
\hline
 & 2 & 150 & 3 & 0.1 & 10 & 1e-04 & 0.25 & 5 & 1e-01 & 100 & 6.0\\
\hline
 & 5 & 150 & 3 & 0.1 & 10 & 1e-04 & 0.25 & 5 & 1e-01 & 100 & 6.0\\
\hline
 & 2 & 100 & 3 & 0.1 & 10 & 1e-01 & 1.00 & 5 & 1e-01 & 25 & 6.0\\
\hline
 & 2 & 100 & 3 & 0.1 & 10 & 1e-01 & 1.00 & 5 & 1e-01 & 33 & 6.0\\
\hline
 & 2 & 100 & 3 & 0.1 & 10 & 1e-01 & 0.50 & 5 & 1e-01 & 33 & 6.0\\
\hline
 & 2 & 100 & 3 & 0.1 & 10 & 1e-01 & 0.50 & 5 & 1e-01 & 33 & 6.0\\
\hline
heart-cleveland & 7 & 50 & 1 & 0.1 & 10 & 1e-01 & 1.00 & 1 & 1e-01 & 40 & 2.0\\
\hline
 & 2 & 50 & 1 & 0.1 & 10 & 1e-01 & 0.50 & 1 & 1e-01 & 40 & 3.5\\
\hline
 & 2 & 50 & 3 & 0.1 & 10 & 1e-04 & 0.50 & 1 & 1e-01 & 40 & 2.0\\
\hline
 & 7 & 100 & 3 & 0.1 & 10 & 1e-01 & 0.50 & 1 & 1e-01 & 25 & 6.0\\
\hline
 & 2 & 50 & 1 & 0.1 & 10 & 1e-04 & 0.25 & 1 & 1e-01 & 25 & 3.5\\
\hline
 & 13 & 100 & 2 & 0.1 & 10 & 1e-01 & 0.25 & 1 & 1e-01 & 100 & 2.0\\
\hline
 & 2 & 50 & 1 & 0.1 & 10 & 1e-04 & 0.50 & 1 & 1e-01 & 40 & 3.5\\
\hline
 & 2 & 50 & 1 & 0.1 & 10 & 1e-04 & 0.50 & 1 & 1e-01 & 40 & 2.0\\
\hline
 & 2 & 50 & 1 & 0.1 & 10 & 1e-01 & 0.25 & 1 & 1e-01 & 100 & 2.0\\
\hline
 & 13 & 100 & 1 & 0.1 & 10 & 1e-04 & 0.25 & 1 & 1e-01 & 100 & 2.0\\
\hline
heart-va & 2 & 150 & 3 & 0.1 & 10 & 1e-04 & 1.00 & 3 & 1e-04 & 40 & 3.5\\
\hline
 & 2 & 50 & 1 & 0.1 & 10 & 1e-04 & 1.00 & 1 & 1e-01 & 100 & 6.0\\
\hline
 & 2 & 100 & 2 & 0.1 & 10 & 1e-04 & 1.00 & 3 & 1e-01 & 100 & 2.0\\
\hline
 & 12 & 100 & 3 & 0.1 & 10 & 1e-01 & 0.50 & 5 & 1e-01 & 25 & 6.0\\
\hline
 & 2 & 100 & 3 & 0.1 & 10 & 1e-04 & 0.25 & 1 & 1e-01 & 100 & 6.0\\
\hline
 & 2 & 50 & 2 & 0.1 & 10 & 1e-04 & 1.00 & 1 & 1e-01 & 40 & 2.0\\
\hline
 & 2 & 100 & 2 & 0.1 & 10 & 1e-04 & 0.50 & 5 & 1e-01 & 100 & 6.0\\
\hline
 & 7 & 50 & 3 & 0.1 & 10 & 1e-04 & 0.50 & 5 & 1e-04 & 100 & 2.0\\
\hline
 & 2 & 100 & 1 & 0.1 & 10 & 1e-01 & 0.50 & 5 & 0e+00 & 100 & 6.0\\
\hline
 & 2 & 50 & 1 & 0.1 & 10 & 1e-01 & 0.25 & 3 & 1e-01 & 100 & 6.0\\
\hline
iris & 2 & 50 & 1 & 0.1 & 10 & 1e-01 & 0.50 & 1 & 1e-04 & 25 & 2.0\\
\hline
 & 3 & 50 & 1 & 0.1 & 10 & 1e-01 & 0.50 & 1 & 1e-01 & 25 & 2.0\\
\hline
 & 2 & 50 & 2 & 0.1 & 10 & 1e-01 & 0.50 & 3 & 1e-01 & 25 & 2.0\\
\hline
 & 2 & 50 & 1 & 0.1 & 10 & 1e-01 & 0.25 & 1 & 1e-04 & 25 & 2.0\\
\hline
 & 2 & 50 & 2 & 0.1 & 10 & 1e-04 & 0.25 & 1 & 1e-04 & 25 & 2.0\\
\hline
 & 2 & 50 & 2 & 0.1 & 10 & 1e-01 & 0.50 & 1 & 1e-01 & 25 & 2.0\\
\hline
 & 2 & 50 & 1 & 0.1 & 10 & 1e-04 & 0.50 & 3 & 1e-01 & 25 & 2.0\\
\hline
 & 2 & 50 & 2 & 0.1 & 10 & 1e-01 & 0.25 & 5 & 1e-04 & 25 & 2.0\\
\hline
 & 2 & 50 & 2 & 0.1 & 10 & 1e-04 & 0.25 & 1 & 1e-04 & 25 & 2.0\\
\hline
 & 2 & 50 & 1 & 0.1 & 10 & 1e-01 & 0.25 & 1 & 1e-04 & 25 & 2.0\\
\hline
lymphography & 2 & 50 & 2 & 0.1 & 10 & 1e-01 & 0.25 & 3 & 0e+00 & 50 & 6.0\\
\hline
 & 2 & 150 & 2 & 0.1 & 10 & 1e-01 & 1.00 & 3 & 1e-01 & 25 & 6.0\\
\hline
 & 2 & 100 & 2 & 0.1 & 10 & 1e-01 & 0.50 & 5 & 1e-01 & 25 & 6.0\\
\hline
 & 2 & 50 & 1 & 0.1 & 10 & 1e-01 & 1.00 & 3 & 1e-01 & 50 & 2.0\\
\hline
 & 2 & 50 & 1 & 0.1 & 10 & 1e-01 & 0.50 & 5 & 1e-01 & 100 & 3.5\\
\hline
 & 2 & 150 & 3 & 0.1 & 10 & 1e-01 & 0.25 & 3 & 0e+00 & 100 & 6.0\\
\hline
 & 2 & 150 & 1 & 0.1 & 10 & 1e-01 & 1.00 & 5 & 1e-01 & 25 & 6.0\\
\hline
 & 2 & 150 & 1 & 0.1 & 10 & 1e-04 & 0.25 & 5 & 0e+00 & 50 & 6.0\\
\hline
 & 2 & 50 & 2 & 0.1 & 10 & 1e-01 & 1.00 & 3 & 1e-01 & 25 & 6.0\\
\hline
 & 2 & 150 & 1 & 0.1 & 10 & 1e-01 & 0.50 & 3 & 1e-01 & 100 & 6.0\\
\hline
pittsburg-bridges-MATERIAL & 4 & 50 & 2 & 0.1 & 10 & 1e-01 & 1.00 & 1 & 1e-01 & 66 & 6.0\\
\hline
 & 2 & 50 & 1 & 0.1 & 10 & 1e-01 & 1.00 & 1 & 1e-01 & 25 & 2.0\\
\hline
 & 2 & 100 & 2 & 0.1 & 10 & 1e-01 & 0.25 & 5 & 1e-01 & 25 & 6.0\\
\hline
 & 2 & 150 & 1 & 0.1 & 10 & 1e-01 & 1.00 & 3 & 1e-04 & 66 & 2.0\\
\hline
 & 4 & 50 & 2 & 0.1 & 10 & 1e-01 & 0.25 & 5 & 0e+00 & 25 & 2.0\\
\hline
 & 4 & 50 & 1 & 0.1 & 10 & 1e-01 & 1.00 & 1 & 1e-01 & 25 & 2.0\\
\hline
 & 2 & 100 & 3 & 0.1 & 10 & 1e-01 & 0.25 & 3 & 1e-04 & 25 & 2.0\\
\hline
 & 2 & 50 & 1 & 0.1 & 10 & 1e-01 & 1.00 & 1 & 1e-01 & 25 & 2.0\\
\hline
 & 2 & 100 & 1 & 0.1 & 10 & 1e-01 & 0.25 & 3 & 1e-01 & 25 & 2.0\\
\hline
 & 4 & 150 & 1 & 0.1 & 10 & 1e-04 & 1.00 & 5 & 0e+00 & 25 & 2.0\\
\hline
pittsburg-bridges-REL-L & 2 & 50 & 3 & 0.1 & 10 & 1e-01 & 0.50 & 1 & 1e-04 & 25 & 2.0\\
\hline
 & 2 & 150 & 1 & 0.1 & 10 & 1e-01 & 0.50 & 5 & 1e-01 & 25 & 3.5\\
\hline
 & 2 & 50 & 1 & 0.1 & 10 & 1e-01 & 0.50 & 1 & 0e+00 & 66 & 6.0\\
\hline
 & 2 & 50 & 2 & 0.1 & 10 & 1e-01 & 1.00 & 1 & 0e+00 & 25 & 3.5\\
\hline
 & 2 & 50 & 1 & 0.1 & 10 & 1e-01 & 0.25 & 1 & 1e-01 & 25 & 2.0\\
\hline
 & 2 & 50 & 2 & 0.1 & 10 & 1e-01 & 0.50 & 5 & 1e-01 & 66 & 2.0\\
\hline
 & 2 & 50 & 2 & 0.1 & 10 & 1e-04 & 0.50 & 5 & 1e-01 & 66 & 6.0\\
\hline
 & 7 & 50 & 1 & 0.1 & 10 & 1e-01 & 0.25 & 5 & 1e-01 & 25 & 6.0\\
\hline
 & 2 & 50 & 1 & 0.1 & 10 & 1e-01 & 0.25 & 1 & 1e-01 & 25 & 3.5\\
\hline
 & 2 & 150 & 1 & 0.1 & 10 & 1e-01 & 1.00 & 5 & 1e-01 & 25 & 6.0\\
\hline
pittsburg-bridges-SPAN & 2 & 100 & 2 & 0.1 & 10 & 1e-01 & 0.25 & 3 & 1e-01 & 66 & 6.0\\
\hline
 & 2 & 150 & 2 & 0.1 & 10 & 1e-01 & 0.25 & 5 & 1e-01 & 66 & 3.5\\
\hline
 & 2 & 50 & 3 & 0.1 & 10 & 1e-01 & 0.50 & 5 & 1e-01 & 25 & 6.0\\
\hline
 & 2 & 100 & 1 & 0.1 & 10 & 1e-01 & 0.50 & 3 & 1e-01 & 25 & 2.0\\
\hline
 & 7 & 100 & 2 & 0.1 & 10 & 1e-01 & 1.00 & 3 & 0e+00 & 100 & 3.5\\
\hline
 & 2 & 150 & 2 & 0.1 & 10 & 1e-01 & 1.00 & 3 & 1e-01 & 25 & 3.5\\
\hline
 & 2 & 100 & 1 & 0.1 & 10 & 1e-01 & 0.25 & 5 & 1e-01 & 66 & 3.5\\
\hline
 & 4 & 50 & 1 & 0.1 & 10 & 1e-01 & 1.00 & 3 & 1e-01 & 66 & 6.0\\
\hline
 & 2 & 50 & 2 & 0.1 & 10 & 1e-01 & 0.25 & 5 & 0e+00 & 100 & 3.5\\
\hline
 & 4 & 100 & 1 & 0.1 & 10 & 1e-01 & 0.50 & 3 & 1e-01 & 25 & 6.0\\
\hline
pittsburg-bridges-TYPE & 2 & 100 & 1 & 0.1 & 10 & 1e-01 & 1.00 & 3 & 1e-01 & 100 & 3.5\\
\hline
 & 2 & 50 & 1 & 0.1 & 10 & 1e-01 & 0.50 & 5 & 1e-01 & 25 & 2.0\\
\hline
 & 2 & 50 & 2 & 0.1 & 10 & 1e-01 & 0.25 & 3 & 1e-01 & 25 & 6.0\\
\hline
 & 2 & 50 & 2 & 0.1 & 10 & 1e-01 & 0.50 & 3 & 1e-01 & 25 & 6.0\\
\hline
 & 2 & 50 & 1 & 0.1 & 10 & 1e-01 & 0.25 & 3 & 1e-01 & 25 & 2.0\\
\hline
 & 2 & 50 & 1 & 0.1 & 10 & 1e-01 & 0.25 & 3 & 1e-01 & 25 & 3.5\\
\hline
 & 2 & 50 & 1 & 0.1 & 10 & 1e-01 & 0.25 & 5 & 0e+00 & 25 & 2.0\\
\hline
 & 2 & 50 & 1 & 0.1 & 10 & 1e-01 & 0.25 & 5 & 1e-04 & 100 & 6.0\\
\hline
 & 2 & 50 & 1 & 0.1 & 10 & 1e-01 & 0.50 & 3 & 1e-01 & 25 & 6.0\\
\hline
 & 2 & 50 & 1 & 0.1 & 10 & 1e-01 & 0.25 & 5 & 1e-01 & 100 & 6.0\\
\hline
seeds & 7 & 150 & 2 & 0.1 & 10 & 1e-04 & 1.00 & 3 & 0e+00 & 25 & 6.0\\
\hline
 & 4 & 50 & 2 & 0.1 & 10 & 0e+00 & 1.00 & 5 & 1e-04 & 100 & 6.0\\
\hline
 & 7 & 100 & 2 & 0.1 & 10 & 0e+00 & 0.50 & 5 & 1e-04 & 66 & 6.0\\
\hline
 & 2 & 100 & 2 & 0.1 & 10 & 1e-04 & 1.00 & 3 & 0e+00 & 25 & 6.0\\
\hline
 & 7 & 100 & 3 & 0.1 & 10 & 0e+00 & 0.50 & 5 & 1e-04 & 25 & 6.0\\
\hline
 & 4 & 100 & 3 & 0.1 & 10 & 1e-04 & 0.25 & 3 & 1e-01 & 25 & 6.0\\
\hline
 & 7 & 100 & 3 & 0.1 & 10 & 1e-04 & 1.00 & 5 & 0e+00 & 100 & 6.0\\
\hline
 & 4 & 100 & 3 & 0.1 & 10 & 1e-04 & 0.50 & 3 & 0e+00 & 100 & 6.0\\
\hline
 & 2 & 50 & 2 & 0.1 & 10 & 0e+00 & 1.00 & 5 & 1e-01 & 25 & 2.0\\
\hline
 & 2 & 150 & 2 & 0.1 & 10 & 1e-04 & 1.00 & 3 & 1e-01 & 25 & 6.0\\
\hline
synthetic-control & 31 & 100 & 3 & 0.1 & 10 & 1e-01 & 0.25 & 5 & 1e-01 & 33 & 3.5\\
\hline
 & 2 & 100 & 2 & 0.1 & 10 & 1e-01 & 0.25 & 5 & 1e-01 & 100 & 6.0\\
\hline
 & 2 & 100 & 3 & 0.1 & 10 & 1e-01 & 0.25 & 5 & 1e-01 & 100 & 6.0\\
\hline
 & 2 & 150 & 3 & 0.1 & 10 & 1e-01 & 0.25 & 5 & 1e-01 & 100 & 6.0\\
\hline
 & 2 & 150 & 2 & 0.1 & 10 & 1e-01 & 0.25 & 5 & 1e-01 & 25 & 6.0\\
\hline
 & 2 & 100 & 2 & 0.1 & 10 & 1e-01 & 0.25 & 5 & 1e-01 & 33 & 6.0\\
\hline
 & 2 & 150 & 2 & 0.1 & 10 & 1e-01 & 0.25 & 5 & 1e-01 & 100 & 6.0\\
\hline
 & 31 & 150 & 2 & 0.1 & 10 & 1e-01 & 0.25 & 3 & 1e-01 & 100 & 3.5\\
\hline
 & 2 & 150 & 2 & 0.1 & 10 & 1e-01 & 0.25 & 5 & 1e-01 & 100 & 6.0\\
\hline
 & 31 & 100 & 3 & 0.1 & 10 & 1e-01 & 0.25 & 5 & 1e-01 & 100 & 6.0\\
\hline
teaching & 5 & 150 & 3 & 0.1 & 10 & 1e-01 & 0.25 & 3 & 1e-04 & 25 & 2.0\\
\hline
 & 2 & 150 & 3 & 0.1 & 10 & 1e-01 & 0.50 & 5 & 1e-01 & 66 & 2.0\\
\hline
 & 5 & 150 & 3 & 0.1 & 10 & 1e-01 & 0.50 & 1 & 1e-04 & 100 & 6.0\\
\hline
 & 2 & 150 & 3 & 0.1 & 10 & 1e-04 & 0.50 & 5 & 1e-04 & 25 & 2.0\\
\hline
 & 3 & 100 & 2 & 0.1 & 10 & 1e-04 & 0.25 & 3 & 1e-01 & 25 & 2.0\\
\hline
 & 2 & 150 & 3 & 0.1 & 10 & 1e-01 & 1.00 & 5 & 1e-01 & 66 & 6.0\\
\hline
 & 2 & 150 & 3 & 0.1 & 10 & 1e-01 & 1.00 & 3 & 1e-04 & 66 & 2.0\\
\hline
 & 2 & 150 & 3 & 0.1 & 10 & 1e-01 & 0.50 & 5 & 1e-01 & 66 & 2.0\\
\hline
 & 5 & 150 & 3 & 0.1 & 10 & 1e-01 & 1.00 & 3 & 1e-01 & 25 & 2.0\\
\hline
 & 2 & 150 & 3 & 0.1 & 10 & 1e-01 & 1.00 & 3 & 0e+00 & 66 & 3.5\\
\hline
vertebral-column-3clases & 4 & 100 & 2 & 0.1 & 10 & 1e-04 & 1.00 & 3 & 1e-01 & 100 & 2.0\\
\hline
 & 2 & 100 & 2 & 0.1 & 10 & 1e-01 & 1.00 & 5 & 1e-04 & 25 & 3.5\\
\hline
 & 2 & 50 & 2 & 0.1 & 10 & 1e-01 & 1.00 & 3 & 1e-01 & 25 & 2.0\\
\hline
 & 2 & 150 & 3 & 0.1 & 10 & 1e-04 & 0.50 & 3 & 1e-01 & 100 & 3.5\\
\hline
 & 2 & 50 & 1 & 0.1 & 10 & 1e-01 & 1.00 & 3 & 1e-01 & 66 & 2.0\\
\hline
 & 2 & 50 & 2 & 0.1 & 10 & 1e-01 & 0.50 & 3 & 1e-01 & 66 & 6.0\\
\hline
 & 2 & 50 & 2 & 0.1 & 10 & 1e-04 & 1.00 & 3 & 1e-01 & 25 & 6.0\\
\hline
 & 2 & 150 & 3 & 0.1 & 10 & 1e-04 & 1.00 & 3 & 1e-01 & 66 & 6.0\\
\hline
 & 2 & 50 & 2 & 0.1 & 10 & 1e-01 & 1.00 & 3 & 1e-01 & 66 & 3.5\\
\hline
 & 2 & 50 & 2 & 0.1 & 10 & 0e+00 & 0.25 & 5 & 1e-01 & 100 & 3.5\\
\hline
wine & 2 & 50 & 2 & 0.1 & 10 & 0e+00 & 0.25 & 5 & 1e-01 & 25 & 2.0\\
\hline
 & 2 & 50 & 3 & 0.1 & 10 & 0e+00 & 0.25 & 3 & 1e-01 & 25 & 2.0\\
\hline
 & 2 & 50 & 2 & 0.1 & 10 & 1e-01 & 0.25 & 3 & 1e-01 & 25 & 3.5\\
\hline
 & 2 & 50 & 1 & 0.1 & 10 & 1e-01 & 0.25 & 3 & 1e-01 & 25 & 2.0\\
\hline
 & 2 & 100 & 2 & 0.1 & 10 & 1e-01 & 0.50 & 3 & 1e-01 & 25 & 3.5\\
\hline
 & 2 & 50 & 3 & 0.1 & 10 & 1e-01 & 0.25 & 3 & 0e+00 & 25 & 2.0\\
\hline
 & 2 & 50 & 1 & 0.1 & 10 & 1e-01 & 0.50 & 3 & 0e+00 & 25 & 2.0\\
\hline
 & 2 & 150 & 3 & 0.1 & 10 & 1e-01 & 1.00 & 5 & 1e-01 & 25 & 3.5\\
\hline
 & 2 & 50 & 2 & 0.1 & 10 & 1e-04 & 0.25 & 3 & 1e-01 & 25 & 2.0\\
\hline
 & 2 & 50 & 1 & 0.1 & 10 & 1e-04 & 0.25 & 5 & 1e-04 & 25 & 2.0\\
\hline
wine-quality-red & 2 & 150 & 3 & 0.1 & 10 & 1e-04 & 1.00 & 1 & 1e-04 & 100 & 6.0\\
\hline
 & 2 & 100 & 3 & 0.1 & 10 & 1e-04 & 1.00 & 3 & 0e+00 & 33 & 6.0\\
\hline
 & 2 & 150 & 2 & 0.1 & 10 & 1e-01 & 1.00 & 3 & 1e-01 & 100 & 3.5\\
\hline
 & 2 & 150 & 3 & 0.1 & 10 & 1e-01 & 1.00 & 3 & 1e-04 & 100 & 2.0\\
\hline
 & 2 & 150 & 3 & 0.1 & 10 & 1e-01 & 1.00 & 1 & 0e+00 & 100 & 6.0\\
\hline
 & 2 & 150 & 3 & 0.1 & 10 & 1e-01 & 1.00 & 1 & 0e+00 & 100 & 3.5\\
\hline
 & 2 & 100 & 3 & 0.1 & 10 & 1e-01 & 1.00 & 3 & 1e-04 & 25 & 6.0\\
\hline
 & 11 & 50 & 3 & 0.1 & 10 & 1e-04 & 1.00 & 3 & 0e+00 & 25 & 6.0\\
\hline
 & 2 & 100 & 2 & 0.1 & 10 & 1e-01 & 1.00 & 3 & 0e+00 & 25 & 2.0\\
\hline
 & 2 & 150 & 3 & 0.1 & 10 & 1e-04 & 1.00 & 3 & 0e+00 & 25 & 2.0\\
\hline

\caption{Parameters chosen by the best CV accuracy on the test set in the UCI datasets, over each of 10 train/validation set splits. For non-BART methods, paramter names above correspond to their name in the \texttt{caret} package, except for ``int depth'' and ``n'' for gbm, which are ``interaction.depth'' and ``n.minobsinnode'' (respectively) in the package. The number of trees in the random forest is fixed at 500 by default in caret.} 
\label{tab:cvpars}

\end{longtable}
}

\subsection{Covariate-dependent heteroscedastic regression}\label{sec:hetmodel}
\newcommand{\vf}{\sigma}
\newcommand{\mv}{m_v}

CGM's BART regression model for continuous data assumed homoscedastic, normally-distributed errors:
\begin{equation}
y_i = \f(\x_i) + \epsilon_i,\quad \epsilon_i\iid N(0,\sigma^2)\label{eq:origianl-bart}\nonumber
\end{equation}
CGM demonstrated that this model performed well relative to competitors in a range of simulations. However, when the data exhibit heteroscedasticity this model may over- or under-fit the mean function, and predictive intervals computed from the posterior predictive $p(y_{n+1}\mid \x_{n+1}, \{y_i, \x_i : 1\leq i\leq n\} )$ will be poorly calibrated. Further, the effect of covariates on the variance may be of interest itself. Heteroscedastic BART models with parametric variance functions were introduced in \cite{Bleich2014}, where the authors provide the necessary expressions for the integrated likelihood and full conditionals to update $\f(\cdot)$ under heteroscedasticity. Here we extend the heteroscedastic BART model to utilize log-linear BART priors (as introduced in the main document) for the variance function. See \citep{Pratola2017-cz} for a detailed exposition of this model under slightly different priors.

Specifically we consider the following regression model:
\begin{equation}
y_i = \f(\x_i) + \vf(\x_i)\epsilon_i,\quad \epsilon_i\iid N(0,\sigma_0^2)\label{eq:hetbart},
\end{equation}
where $\vf^{2}(\cdot)$ is given a log-linear BART prior:
$$
\log[\sigma^2(\x)] = \sum_{h=1}^m \g(\x, T^{\sigma}_h, \M^{\sigma}_h).
$$

Due to the symmetry of our prior distribution this is exactly equivalent to a log-linear BART prior on $\vf^{-2}(\cdot)$, or a log-linear BART prior on the standard deviation function $\sigma(\x)$ where the prior standard deviation on the leaf parameters is scaled by half. We give the mean function $\f(\cdot)$ a BART prior with normal priors on the end node parameters as in the original BART prior. Rather than centering and scaling $y$ to $\pm 0.5$, we scale $\sigma_\mu$ by $0.5(y_{max} - y_{min})$ which has much the same effect. The parameter $\sigma^2_0$ can be formally elicited or chosen using a slight adaptation of CGM's heuristic for setting the scale parameter in the prior on the error variance in homoskedastic BART.
Smaller values of $a_0$ tend to be necessary to avoid overfitting. Taking $a_0=1.5$ ensures that the marginal prior for the variance function $1/\vf(\x_i)$ puts approximately 95\% prior probability on $\sigma^2_0\f_v(\x)\in(\sigma^2_0/5, 5\sigma^2_0)$.

\subsubsection{MCMC}
The likelihood for a single data point is
\beq
p(y_i)=
\frac{\vf(\x_i)^{-1}}{\sqrt{2\pi\sigma_0^2}}\exp\left[-\frac{1}{2\vf(\x_i)\sigma_0^2}\left(y_i - \f(\x_i)\right)^2 \right].
\eeq
The log-linear BART prior is immediately conjugate, so no data augmentation is necessary. For updating the trees and parameters in $\vf(\cdot)$, $\f(\cdot)$ is considered fixed. MCMC in the heteroscedastic model proceeds as follows:

\begin{enumerate}
\item Update the mean function's trees and node parameters as in \cite{Bleich2014}, using $\sigma^2_0\vf(\x_i)$ as the variance for each observation.
\item %
Update $\sigma^2(\cdot)$ using Algorithm~\ref{alg:backfit} and the expressions in Section~\ref{sec:mcmc}, with
\begin{gather}
r_{ht} = \frac{1}{2}\sum_{i=1}^n \ind{\x_i\in \mathcal{A}^{(v)}_{ht}}\\
s_{ht} = \frac{1}{2\sigma_0^2}\sum_{i:\x_i\in \mathcal{A}^{(v)}_{ht}} \noth(\x_i)(y_i - \f(\x_i))^2
\end{gather}

\end{enumerate}

\bibliographystyle{./plainnat}
\bibliography{refs}